%% file: organoid_R1.tex
\documentclass[a4paper]{article} 
\usepackage{lipsum}
\usepackage{url}
\usepackage{mathpazo}
\usepackage[T1]{fontenc}
\usepackage[utf8]{inputenc}
\usepackage[english]{babel}

\usepackage[left=3cm,top=3cm,right=3cm,bottom=3cm]{geometry}

\usepackage{natbib}
\usepackage{amsmath}
\usepackage{amssymb}
\usepackage{amsthm}
\usepackage{hyperref}
\usepackage{graphicx}
\usepackage{longtable}
\usepackage{verbatimbox}

\usepackage[title]{appendix}
\usepackage{bm}
\usepackage{placeins}

\usepackage{subcaption}
\usepackage[small, sc]{titlesec}

\usepackage{xcolor}
\hypersetup{
colorlinks,
linkcolor={red!50!black},
citecolor={blue!50!black},
urlcolor={blue!80!black}
}
\input{defs}

\begin{document}
\title{\textsc{Surface tension controls the onset of gyrification in brain organoids}}

\author{\textsc{Davide Riccobelli}$^1$\thanks{\href{mailto:davide.riccobelli@sissa.it}{\texttt{davide.riccobelli@sissa.it}}} $\,\,\cdot$
\textsc{Giulia Bevilacqua}$^{2}$\thanks{\href{mailto:giulia.bevilacqua@polimi.it}{\texttt{giulia.bevilacqua@polimi.it}}}\bigskip\\
\normalsize$^1$ SISSA -- International School for Advanced Studies, via Bonomea 265, Trieste, Italy.\\
\normalsize$^2$ MOX -- Politecnico di Milano, piazza Leonardo da Vinci 32, Milano, Italy}
\date{\today}

\maketitle

\begin{abstract}
Understanding the mechanics of brain embryogenesis can provide insights on pathologies related to brain development, such as \emph{lissencephaly}, a genetic disease which causes a reduction of the number of cerebral sulci. Recent experiments on brain organoids have confirmed that gyrification, i.e.~the formation of the folded structures of the brain, is triggered by the inhomogeneous growth of the peripheral region. However, the rheology of these cellular aggregates and the mechanics of lissencephaly are still matter of debate.\\
In this work, we develop a mathematical model of brain organoids based on the theory of morpho-elasticity. We describe them as non-linear elastic bodies, composed of a disk surrounded by a growing layer called cortex. The external boundary is subjected to a tissue surface tension due the intercellular adhesion forces. We show that the resulting surface energy is relevant at the small length scales of brain organoids and affects the mechanics of cellular aggregates. We perform a linear stability analysis of the radially symmetric configuration and we study the post-buckling behaviour through finite element simulations.\\
We find that the process of gyrification is triggered by the cortex growth and modulated by the competition between two length scales: the radius of the organoid and the capillary length generated by surface tension. We show that a solid model can reproduce the results of the \emph{in-vitro} experiments. Furthermore, we prove that the lack of brain sulci in lissencephaly is caused by a reduction of the cell stiffness: the softening of the organoid strengthens the role of surface tension, delaying or even inhibiting the onset of a mechanical instability at the free boundary.
\end{abstract}

\section{Introduction}

The formation of folded structures in human and animal brains makes it possible to increase the extension of the cerebral cortex, packing a larger number of neurons in a limited space. The creation of these furrows and ridges called \emph{sulci} and \emph{gyri}, respectively, is fundamental for a healthy development of the brain in embryogenesis. The mechanics underlying this morphogenetic phenomenon is not still completely understood.\\
Recent experiments performed on human brain organoids \citep{Karzbrun_2018}, i.e. cell agglomerates cultured in vitro that reproduce the morphogenesis of organs, apparently confirm the hypothesis that sulci are generated by brain cortex buckling triggered by growth \citep{Ronan_2013, budday2014role, Bayly_2014}. In \citep{Karzbrun_2018}, the authors observed an increased growth of the cortex with respect to the underlying lumen. 
In some pathological situations such as \emph{lissencephaly}, a genetic mutation, the physiological generation of brain sulci is inhibited or even suppressed. This disease, caused by the LIS1 heterozygous ($+/-$) mutation, is correlated to nutritional disorders, alterations in muscle tone, severe psychomotor and mental retardation \citep{Dobyns_1993}. The mathematical description of brain sulci embryogenesis can provide new insights to understand the mechanisms underlying this disease.

A well developed framework to model mechanical instabilities induced by growth is the theory of \emph{morpho-elasticity} \citep{Goriely_2017}, where living tissues are treated as growing elastic materials. A spatially inhomogeneous growth generates microstructural misfits, leading to a geometrically incompatible relaxed configuration. The restoration of the compatibility requires elastically distorting the body and generating residual stresses \citep{Hoger_1985, Rodriguez_1994}.\\
Differential growth and residual stresses are involved in the morphogenesis of tissues such as intestinal villi \citep{Balbi_2013, Ben_Amar_2013, Ciarletta_2014} and they enhance the mechanical strength of several biological structures, such as arteries \citep{Chuong_1986}.

A first model of the experiments on brain organoids \citep{Karzbrun_2018} has been developed by \cite{balbi2018mechanics}. The authors model the organoid as a non-linear elastic material, where gyrification is triggered by a remodelling of the cortex and the contraction of the lumen. In their model, the selection of the critical wavelength is dictated by different mechanical properties of the lumen and the cortex. Despite the good agreement with experimental results, \cite{Engstrom_2018} noticed that brain organoids exhibit an unconventional behaviour: the cortex is thinner in correspondence of sulci and thicker in correspondence of gyri, in contrast with the morphology predicted by elastic models. 

\begin{figure}[t!]
\centering
\includegraphics[width=0.8\textwidth]{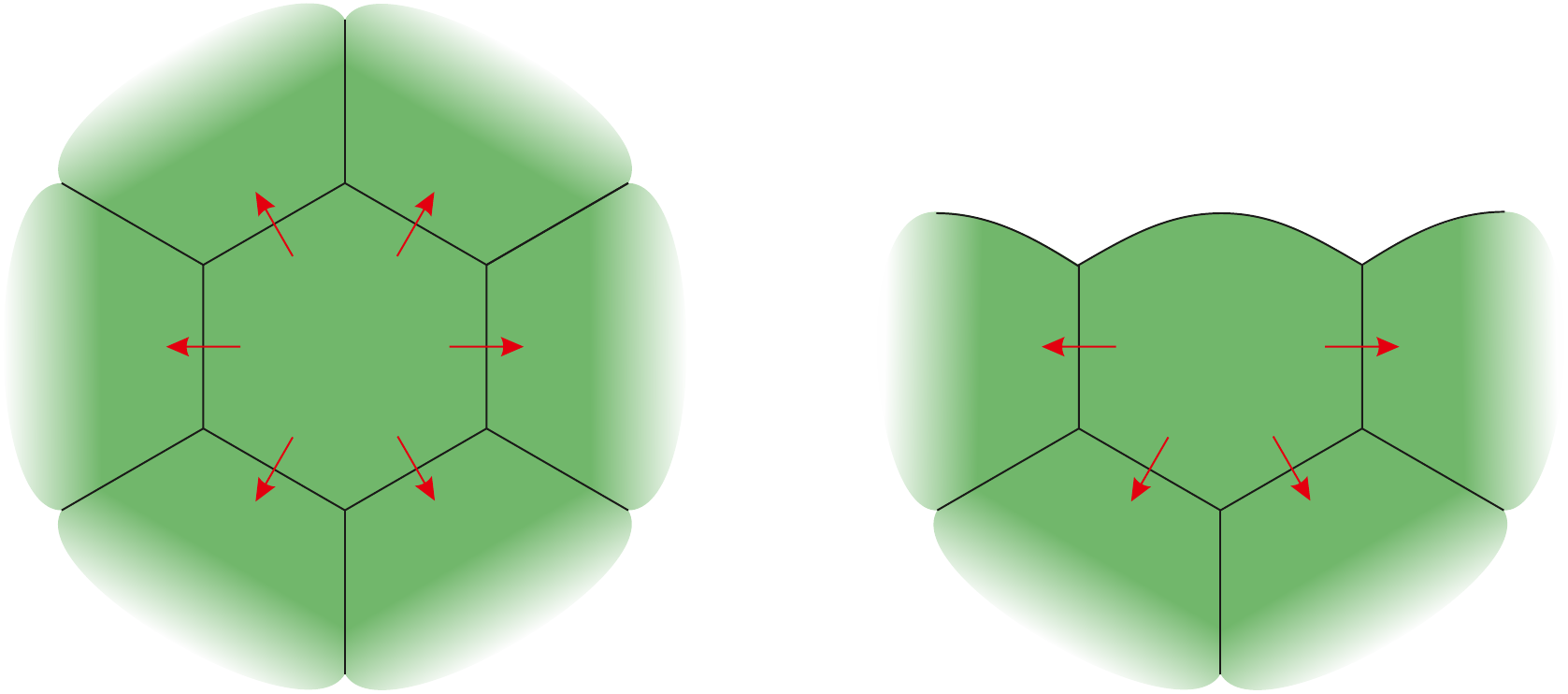}
\caption{Cells in the bulk (left) and on the free surface (right) of a cellular lattice. Adhesion forces generated by the surrounding cells are denoted by the red arrows. The sum of all these forces on an internal cell is zero, while it is non-zero and perpendicular to the boundary for a cell on the free surface.}
\label{fig:surf_tens}
\end{figure}

In this paper, we revisit the model proposed by \cite{balbi2018mechanics} to overcome the limitations remarked by \cite{Engstrom_2018}, proposing a different explanation of lissencephaly.\\
In cellular aggregates, cohesion among cells is due to adhesion forces induced by adhesion mole\-cules \citep{Turlier_2015, Ma_tre_2015}. Internal cells are surrounded by other cells, so that the sum of all these forces is zero and each cell is in mechanical equilibrium. Conversely, cells at the boundary of the agglomerate possess a portion of their membrane which is not in contact with other cells: the total adhesion force acting on such cells is non null and it is perpendicular to the free surface of the cellular agglomerate (see Fig.~\ref{fig:surf_tens} for a graphical representation). These forces generate deformation and the appearance of a boundary layer at the periphery that can be treated as a surface effect called \emph{tissue surface tension} \citep{Steinberg_1963}.

This phenomenon is reminiscent of the mechanics of surface tension in fluids and soft gels \citep{Style_2017}. Since organoids and embryos are characterized by small length scales, surface effects arising from cohesion forces cannot be neglected. The presence of tissue surface tension has been used in fluid models of cellular agglomerates \citep{foty1996surface, Davis_1997, Forgacs_1998} but it is usually overlooked in solid models.

Surface tension has been used within the framework of the theory of non-linear elasticity to regularize the \cite{Biot} instability, i.e. the buckling instability of an elastic half-space under compression. Without taking into account surface tension, all the wrinkling modes become unstable at the same compression rate \citep{Ben_Amar_2010}. Non-linear elasticity and particularly models based on elastic bilayers \citep{Li_2012, Cao_2012, Jia_2015, Jin_2019} have been widely employed to model growth phenomena in living tissues, such as the morphogenesis of intestinal villi \citep{Ben_Amar_2013, Ciarletta_2014} and of brain sulci \citep{Tallinen_2014, Holland_2018}.
The majority of the literature (see for instance \citep{jia2018curvature, Balbi_2019} for solid cylindrical bi-layers) focuses their attention on growth induced instability without considering surface tension. A remarkable exception is provided by the work of \cite{Dervaux_2011}, where surface tension is introduced in a cylindrical bi-layer as a regularization of the Biot instability: when the substrate is much more rigid than the coating, the critical wavelength goes to zero. A possible method to regularize this instability is to introduce a small surface tension that introduces a new length scale (the capillary length), leading to a finite critical wavelength (see \cite{Ben_Amar_2010}).
At small length scales, elasto-capillary forces can deform soft gel beams \citep{Mora_2010, Mora_2013} even inducing mechanical instabilities \citep{Taffetani_2015, Xuan_2016}. Surface tension can also enhance the resistance to fracture in soft solids~\citep{Liu_2014, Hui_2016}.

The work is organized as follows: in Section~\ref{sec:surface_tension} we justify the assumption of a solid model for brain organoids and we compute an estimation of the surface tension acting on a solid cellular aggregate. In Section~\ref{sec:elastic_model} we develop the elastic model of brain organoids. In Section~\ref{sec:lin_stab} we perform a linear stability analysis of the radially symmetric configuration and in Section~\ref{sec:post_buckling} we implement a finite element code to study the post-buckling behaviour. Finally, in Section~\ref{sec:discussion} we discuss the outcomes of our model together with some concluding remarks.

\section{Intercellular adhesion generates surface tension in cellular aggregates}
\label{sec:surface_tension}


At the micro-scale, cellular aggregates are composed by several constituents that, in bulk, have a solid or a fluid-like mechanical nature, like cells, the extracellular matrix, the interstitial fluid. From a macroscopic point of view, these agglomerates can be treated as continuum media but their rheology is still a matter of debate. In fact, cellular aggregates are frequently modelled as fluids \citep{Foty_1994, Manning_2010} that can bear external loading thanks to the tissue surface tension. Another point of view is that cellular aggregates behave as active viscoelastic solids \citep{Kuznetsova_2007,Ambrosi_2017, Karzbrun_2018}.

Some important features of biological tissues mechanics cannot be reproduced by fluid-like models. First, there are experimental evidences that cell mitosis and apoptosis (i.e. the cell division and death respectively) are regulated by mechanical stress \citep{Montel_2012, Montel_2012_2}. In particular, in \citep{Cheng_2009}, the authors report an increased cell duplication in the regions where the compressive stress exerted by the surrounding material on the tumour spheroid is minimum. This spatially inhomogeneous growth can be explained only by using a solid description of the cellular agglomerate: the stress tensor of a fluid at rest corresponds to an hydrostatic pressure which is independent on the spatial position \citep{Ambrosi_2017}.

Furthermore, contrarily to fluids, solids can store mechanical stress even in the absence of external loads. These stresses are called residual \citep{Hoger_1985} and they are created when differential growth in solid matter develops microstructural misfits. These geometrical incompatibilities are restored by elastic distortions of the body, generating mechanical stress \citep{Rodriguez_1994, Goriely_2017, Riccobelli_2019}.

Contrary to the fluid approach, the tissue surface tension of cellular aggregates is frequently neglected in solid models. Nonetheless, solids  possess surface tension too \citep{Style_2017}: it can play an important role, when the aggregate is very soft or has a small size. In fact, surface tension introduces a new length scale in the problem: let $\mu$ be the shear modulus of the cellular agglomerate and $\gamma$ the surface tension, then the capillary length $\ell_\text{c}$ is defined as \citep{Mora_2013, Style_2017}
\[
\ell_\text{c} = \frac{\gamma}{\mu}.
\]

Whenever this length scale is of the same order as the characteristic length of the body (e.g. the radius of a spheroid) surface tension cannot be neglected: it can produce a non-negligible deformation \citep{Mora_2013, Style_2013, Mora_2015} and it can even induce mechanical instabilities \citep{Mora_2010, Taffetani_2015, Xuan_2016}. Cellular aggregates are very soft and the effect of the surface tension can be highly relevant, as we show in the following.

\subsection{Estimation of the surface tension acting on a multicellular spheroid}
\label{sec:surf_tens_spheroid}

There are experimental evidences that a tensile skin, having the thickness of a couple of cells, generates an isotropic compression inside multicellular aggregates \citep{Lee_2019}. This phenomenon can be explained as a manifestation of tissue surface tension induced by intercellular adhesion: the tensile skin is indeed very thin and such boundary layer can be treated as a surface tension.

Modelling the unloaded multicellular spheroid as a ball occupying the domain
\[
\Omega_\text{s} = \left\{\vect{X}\in\R^3\;|\;|\vect{X}|<R_\text{o}\right\},
\]
we assume that the spheroid is composed of an incompressible elastic material. Let $\tens{T}$ be the Cauchy stress tensor, the balance of the linear and angular momentum reads
\begin{equation}
\label{eq:equi_sphero}
\diver \tens{T} = \vect{0}.
\end{equation}
If the spheroid is subjected to a tissue surface tension $\gamma$, the boundary condition reads \citep{Style_2017}
\begin{equation}
\label{eq:surf_tens_spheroid}
\tens{T}\vect{n} = \gamma \mathcal{K} \vect{n}, \qquad\text{at }R = R_\text{o}
\end{equation}
where $R$ denotes the radial position, $\mathcal{K}$ is twice the mean curvature, and $\vect{n}$ is the external normal.

We now show that the undeformed reference configuration is in mechanical equilibrium. Experimental observations \citep{Lee_2019} lead us to state that residual stresses are absent. The Cauchy stress is then given by \citep{ogden1997non}
\[
\tens{T} = -p\tens{I}
\]
where $\tens{I}$ is the identity and $p$ is the pressure field that enforces the incompressibility constraint. A constant pressure field $p$ satisfies the equilibrium equation \eqref{eq:equi_sphero}. By imposing Eq.~\eqref{eq:surf_tens_spheroid}, we get
\[
p = \frac{2 \gamma}{R_\text{o}}
\]
which is nothing but the Young-Laplace equation. We invert the previous equation with respect to $\gamma$, obtaining
\begin{equation}
\label{eq:surf_tens_spheroid_2}
\gamma = \frac{p R_\text{o}}{2}.
\end{equation}
From the work of \cite{Lee_2019}, we get that the typical radius of a spheroid is $\sim 400\,\mu\mathrm{m}$ and the internal pressure $p$ is about $500\,\text{Pa}$ (Fig.~5d in \citep{Lee_2019}). From these data and from Eq.~\eqref{eq:surf_tens_spheroid_2}, we estimate that the surface tension acting on the spheroid is $\gamma\simeq 0.1\,\mathrm{N}/\mathrm{m}$.

\begin{rem}
In this Section, we have estimated the tissue surface tension acting on a spheroid. Unfortunately, up to our knowledge, there are not similar experimental data that can be used to perform the same computation for brain organoids.
There are some measurements of the surface tension acting on different embryonic tissues when they are treated as fluids \citep{Sch_tz_2008}. In their work, all the measured values have the same order of magnitude. From these data, we can assume that $\gamma\simeq 0.1\,\mathrm{N}/\mathrm{m}$ is a qualitative estimate of tissue surface tension acting on a brain organoid.
\end{rem}

\section{Elastic model of brain organoids}
\label{sec:elastic_model}

In this Section, we illustrate a model of brain organoids, described as growing hyperelastic bodies subjected to surface tension.

\subsection{Kinematics}

We denote by $\vect{X}$ the material position coordinate. Since brain organoids are cell aggregates confined in a narrow space by a coverslip and a membrane \citep{Karzbrun_2018}, we model them as two dimensional objects. Let
\[
\Omega_0 = \left\{\vect{X} = [R \cos\Theta,\,R\sin\Theta]\in\R^2 \;|\;0\leq R<R_\text{o}\text{ and }0\leq\Theta<2\pi\right\}
\]
be the reference configuration of the organoid. We indicate with $\vect{\varphi}:\Omega_0\rightarrow\R^2$ the deformation field, so that the actual configuration of the body $\Omega$ is given by $\vect{\varphi}(\Omega_0)$. Let $\vect{x}= \vect{\varphi}(\vect{X})$ be the actual position of the point $\vect{X}$ and the displacement vector is defined as $\vect{u}(\vect{X}) = \vect{\varphi}(\vect{X})-\vect{X}$. Let $\tens{F}$ be the deformation gradient, i.e. $\tens{F} = \Grad\vect{\varphi}$.
We exploit a multiplicative decomposition of the deformation gradient (known as Kr\"oner-Lee decomposition \citep{Kr_ner_1959,Lee_1969}) to model the growth of the organoid, so that
\[
\tens{F} = \tens{F}_\text{e}\tens{G}
\]
where the growth tensor $\tens{G}$ accounts for the local inelastic distortion due to the body growth, while $\tens{F}_\text{e}$ describes the elastic distortion necessary to maintain the geometrical compatibility of the body and to balance the external and internal forces \citep{Rodriguez_1994}. As commonly done for biological tissues, it is reasonable to model organoids as incompressible media, namely we enforce that
\begin{equation}
\label{eq:incompressibility}
\det\tens{F}_\text{e} = 1.
\end{equation}

In \citep{Karzbrun_2018}, the authors identified two distinct regions in brain organoids: an internal lumen and an external ring, called cortex, the latter being characterized by a faster growth. Indicating with $R_\text{i}$ the radius of the lumen, we denote these two regions by $\Omega_{0\text{L}}$ and $\Omega_{0\text{R}}$:
\[
\Omega_{0\text{L}} = \{\vect{X}\in\Omega\;|\;R<R_\text{i}\},\qquad
\Omega_{0\text{C}} = \{\vect{X}\in\Omega\;|\;R_\text{i}< R <R_\text{o}\},
\]
and their images through $\vect{\varphi}$ are denoted by $\Omega_\text{L}$ and $\Omega_\text{R}$ respectively.

We assume that the growth tensor $\tens{G}$ takes diagonal form
\begin{equation}
\label{eq:tensG}
\tens{G} = \left\{
\begin{aligned}
&\tens{I} &&\text{if }R<R_\text{i},\\
&g\tens{I} &&\text{if }R_\text{i}< R < R_\text{o},
\end{aligned}
\right.
\end{equation}
where the scalar quantity $g$ is the growth rate of the cortex with respect to the lumen. Tissue surface tension acts at both the interfaces organoid-coverslip and organoid-membrane. Let us call $\gamma_\mathrm{membrane}$ and $\gamma_\mathrm{coverslip}$ the tension acting on the organoid at the interface with these two surfaces. Their contribution to the total energy is given by
\[
E_\mathrm{interface}=\int_\Omega [\gamma_\mathrm{membrane}+\gamma_\mathrm{coverslip}]\,dv = (\gamma_\mathrm{membrane}+\gamma_\mathrm{coverslip})|\Omega_0|\det\tens{F}.
\]
Since the deformation is isochoric, $\det\tens{F} = \det\tens{G}$. The contribution of the surface tension acting on both the membrane and the coverslip, given by $E_\mathrm{interface}$, does not change when the organoid undergoes an elastic deformation and it does not influence either the stability or the morphology of the organoid.
This agrees with the observations of \cite{Karzbrun_2018}: the authors tried different coverslips and membranes, each one characterized by a different adhesion to the organoid and they did not notice remarkable differences (see Fig.~S9 of the supplementary material of \cite{Karzbrun_2018}). From now on, since $E_\text{interface}$ does not influence the deformation of the body, it is not explicitly included in the total energy.\\
We now introduce some mechanical constitutive assumptions.

\subsection{Mechanical constitutive assumptions and force balance equations}
We assume that the organoids are composed of a homogenous hyperelastic material, having strain energy density $\psi$. The first Piola--Kirchhoff stress $\tens{P}$ and the Cauchy stress tensors $\tens{T}$ are then given by
\[
\tens{P} = (\det{\tens{G}})\tens{G}^{-1}\frac{\partial\psi_0(\tens{F}_\text{e})}{\partial\tens{F}_\text{e}}-p\tens{F}^{-1}\qquad\tens{T} = \frac{1}{\det\tens{F}}\tens{F}\tens{P}
\] 
where $p$ is the Lagrange multiplier enforcing the incompressibility constraint \eqref{eq:incompressibility}. The balance of the linear and angular momentum reads
\begin{equation}
\label{eq:balance}
\Diver\tens{P} = \vect{0}\text{ in }\Omega_{0\text{L}},\,\Omega_{0\text{C}},\qquad\text{or}\qquad\diver\tens{T}=\vect{0}\text{ in }\Omega_\text{L},\,\Omega_\text{C}
\end{equation}
in the material and actual reference frame, respectively.

We assume that the center of the organoid is fixed, i.e.
\begin{equation}
\label{eq:BCcenter}
\vect{u}(\vect{0}) = \vect{0} 
\end{equation}
while a constant surface tension $\gamma$ acts at the external boundary of the organoid, so that \citep{Style_2017}
\begin{equation}
\label{eq:BC_capi_T}
\tens{T}\vect{n} = \gamma \mathcal{K} \vect{n}
\end{equation}
where $\vect{n}$ is the outer normal in spatial coordinates and $\mathcal{K}$ is the oriented curvature of the boundary curve $\vect{\mathcal{L}}$ parametrized clockwise, i.e.
\begin{equation}
\label{eq:clockwise}
\vect{\mathcal{L}}(\Theta)  = \vect{\varphi}\left([R_o\cos(\Theta),\,R_o\sin(-\Theta)]\right).
\end{equation}
The Lagrangian form of the boundary condition \eqref{eq:BC_capi_T} is obtained performing a pull-back
\begin{equation}
\label{eq:PN}
\tens{P}^T\vect{N} = (\det{\tens{F}})\gamma \mathcal{K} \tens{F}^{-T}\vect{N}.
\end{equation}
Finally, we enforce the continuity of the stress at the interface $R = R_\text{i}$, so that
\begin{equation}
\label{eq:cont_stress}
\lim_{R\rightarrow R_\text{i}^-}\tens{P}^T\vect{N} = \lim_{R\rightarrow R_\text{i}^+}\tens{P}^T\vect{N}.
\end{equation}
We can assume that brain organoids behave as isotropic media, since they are composed of neural progenitors and not fully developed neurons \citep{Karzbrun_2018}. Moreover, in the fully developed brain, both the gray and the white matter are essentially isotropic \citep{Budday_2017, Budday_2017b}. As it concerns the internal structure, 
brain organoids are complex cell agglomerates composed of different biological materials, such as the extra-cellular matrix, cells, interstitial fluid. According to the observations reported in \cite{Karzbrun_2018}, the extra cellular matrix and the cellular cytoskeleton account for the elasticity of the organoid.
We have assumed that the two elastic phases in the organoids generate a homogeneous hyperelastic material.
There are several works which show a very good fit of the experimental data by assuming a Mooney--Rivlin model (especially when the stretches are moderate \citep{Destrade_2015, Balbi_2019}). Such an energy reduces to the neo-Hookean one whenever the deformation is planar. Hence, the strain energy density is given by
\begin{equation}
\label{eq:strain_energy}
\psi(\tens{F}) =\det(\tens{G})\psi_0(\tens{F}_\text{e}) =(\det\tens{G})\frac{\mu}{2}\left(\tr(\tens{F}_\text{e}^T\tens{F}_\text{e})-2\right).
\end{equation}
The first Piola--Kirchhoff and Cauchy stress tensors read, respectively,
\begin{equation}
\label{eq:stress_tensors}
\left\{
\begin{aligned}
&\tens{P} = \mu (\det{\tens G}) \tens{G}^{-1}\tens{G}^{-T}\tens{F}^{-T} - p\tens{F}^{-1},\\
&\tens{T} = \mu \tens{F}\tens{G}^{-1}\tens{G}^{-T}\tens{F}^T - p\tens{I}.
\end{aligned}
\right.
\end{equation}

Summing up, Eqs.~\eqref{eq:incompressibility} and \eqref{eq:balance}, together with the kinematic constraint Eq.~\eqref{eq:BCcenter} and the boundary condition Eq.~\eqref{eq:PN} define the non linear elastic problem. In the next Section we look for a radially symmetric solution.

\subsection{Equilibrium radially-symmetric solution}
Let $(r,\,\theta)$ be the actual radial and polar coordinates of a point. Let $(\vect{E}_R,\,\vect{E}_\Theta)$ and $(\vect{e}_r,\,\vect{e}_\theta)$ be the local vector basis in polar coordinates in the Lagrangian and Eulerian reference frame, respectively. We look for a radially-symmetric solution of the form
\[
\vect{\varphi}(\vect{X}) = r(R)\vect{e}_r.
\]
The deformation gradient expressed in polar coordinates reads
\begin{equation}
\label{eq:Fbase}
\tens{F} = 
\diag\left(r', \frac{r}{R}\right).
\end{equation}
It is immediate to notice that
\[
r(R) =  R\qquad\text{for }R<R_\text{i},
\]
where $r_\text{i} = r(R_\text{i}) = R_\text{i}$. In the cortex, from the incompressibility constraint given by Eq.~\eqref{eq:incompressibility}, we get
\[
r'r = g^2R.
\]
Performing an integration and imposing that $R_\text{i}=r_\text{i}$, we get
\begin{equation}
\label{eq:r}
r(R) = g\sqrt{R^2 +\left(\frac{1}{g^2}-1\right)R_\text{i}^2}.
\end{equation}

It remains to determine the pressure field $p$. First, we notice that, inverting and differentiating \eqref{eq:r}, we obtain
\begin{equation}
\label{eq:Rr'}
\left\{
\begin{aligned}
&R = \frac{1}{g}\sqrt{r^2+(g^2-1)R_\text{i}^2},\\
&r' = g\frac{\sqrt{r^2+(g^2-1)R_\text{i}^2}}{r},
\end{aligned}
\right.\qquad\text{in }R_\text{i}<R<R_\text{o}
\end{equation}
respectively. The curvature of the boundary line is $-r_\text{o}^{-1}$, where
\[
r_\text{o} = r(R_\text{o}) = g\sqrt{R_\text{o}^2 +\left(\frac{1}{g^2}-1\right)R_\text{i}^2}.
\]
The boundary condition Eq.~\eqref{eq:BC_capi_T} reads
\[
\tens{T}\vect{e}_r = -\frac{\gamma}{r_\text{o}}\vect{e}_r.
\]
Since the deformation depends only on the radial position $r$, the balance of the linear and angular momentum in polar coordinates reads
\begin{equation}
\label{eq:divTrad}
\frac{d T_{rr}}{dr} + \frac{T_{rr}-T_{\theta\theta}}{r}=0
\end{equation}
where $T_{ij}$ are the components of the Cauchy stress tensor $\tens{T}$ in polar coordinates. From Eqs.~\eqref{eq:stress_tensors}, \eqref{eq:Fbase} and \eqref{eq:Rr'}, the Cauchy stress in the cortex reads
\begin{equation}
\label{eq:Tcortex}
\tens{T} = \diag\left(
\frac{\mu  \left(R_\text{i}^2 \left(g^2-1\right)+r^2\right)}{r^2}-p ,\,\frac{\mu  r^2}{R_\text{i}^2 \left(g^2-1\right)+r^2}-p\right).
\end{equation}
We can integrate Eq.~\eqref{eq:divTrad} from $r$ to $r_\text{o}$, obtaining
\begin{equation}
\label{eq:Trrcortex}
T_{rr}(r) = -\frac{\gamma}{r_\text{o}}+\int^{r_\text{o}}_r\left[\frac{\mu  \left(-\frac{\rho ^4}{R_\text{i}^2 \left(g^2-1\right)+\rho ^2}+R_\text{i}^2 \left(g^2-1\right)+\rho ^2\right)}{\rho ^3}\right]\,d\rho.
\end{equation}
We can find the pressure field in the cortex (i.e. for $r_\text{i}<r<r_\text{o}$) by plugging Eq.~\eqref{eq:Tcortex} into Eq.~\eqref{eq:Trrcortex}, obtaining
\begin{equation}
\label{eq:pCortex}
\begin{multlined}
p = f_p(r) := \frac{1}{2} \left(\mu  \left(\frac{R_{\text i}^2 \left(g^2-1\right)}{r^2}+2\right)+\right.\\
+\mu  \left(-\log \left(R_{\text i}^2 \left(g^2-1\right)+r^2\right)+\log \left (R_{\text i}^2 \left(g^2-1\right)+r_{\text o}^2\right)+2 \log \left(\frac{r}{r_{\text o}}\right)\right)+\\
\left.+\frac{\mu  R_{\text i}^2 \left(g^2-1\right)+2 \gamma  r_{\text o}}{r_{\text o}^2}\right).
\end{multlined}
\end{equation}
Finally, we impose the continuity of the stress Eq.~\eqref{eq:cont_stress} at $r = r_\text{i} = R_\text{i}$ to get the pressure for $r<r_\text{i}$. Since the lumen remains undeformed, the Cauchy stress reads
\[
\tens{T} = -p_\text{L}\tens{I}.
\]
Using Eq.~\eqref{eq:Tcortex} and Eq.~\eqref{eq:pCortex}, we can write the solution
\begin{align}
\label{eq:r<ri}
&\left\{
\begin{aligned}
&r =  R\\
&p = p_\text{L} :=  f_p(r_\text{i})+\mu -g^2 \mu
\end{aligned}
\right.&&\text{for }r<r_\text{i},\\
\label{eq:ri<r<ro}
&\left\{
\begin{aligned}
&r = g\sqrt{R^2 +\left(\frac{1}{g^2}-1\right)R_\text{i}^2}\\
&p = p_\text{C} :=  f_p(r)
\end{aligned}
\right.&&\text{for }r_\text{i}<r<r_\text{o}.
\end{align}
In the next section a linear stability analysis of the solution given by Eqs.~\eqref{eq:r<ri}-\eqref{eq:ri<r<ro} is performed, but first we remark some important aspects regarding the boundary conditions.

\begin{rem}
In this model, we have assumed that the only force acting on the boundary of the organoid is due to tissue surface tension. We remark that, in the experiments reported in \citep{Karzbrun_2018}, organoids are embedded in Matrigel. Karzbrun and co-authors tried different concentration of the gel, without registering remarkable changes to the morphology of the wrinkling (see Fig.~S9 \citep{Karzbrun_2018}), so that we are lead to conjecture that Matrigel has a negligible effect on the onset of the instability.

Matrigel is characterized by an elastic modulus of the same order as the organoid one ($500$ Pa \cite{Soofi_2009}). In each experiment, eleven organoids are placed in a dish having a diameter of 6 cm, so that the distance among organoids is of the order of centimeters.
The distance among organoids is much larger than the displacement induced by the organoid growth, so that the response of the Matrigel is linear and can be modelled as the action of linear springs acting at the boundary, as done for similar motivations in \citep{Riccobelli_2018}. We have verified through this approach that the role of Matrigel is negligible.
For the readers interested in the modeling of the Matrigel embedment, we expose in Appendix \ref{app:B} some details on the linear stability analysis in the presence of linear springs at the boundary.
\end{rem}
\section{Linear stability analysis}
\label{sec:lin_stab}

\subsection{Incremental equations}
We apply the theory of incremental deformations superposed on finite strains \citep{ogden1997non} to investigate the stability of the radially symmetric solution. Let $\delta \vect{u}$ be the incremental displacement field and let $\tens{\Gamma} = \grad\delta \vect{u}$. We introduce the push-forward of the incremental Piola-Kirchhoff stress in the axisymmetric deformed configuration, given by
\begin{equation}
\label{eq:A0}
\delta \tens{P} = \mathcal{A}_0 : \tens{\Gamma} +p \tens{\Gamma} - \delta p \tens{I}
\end{equation}
where $\mathcal{A}_0$ is the fourth order tensor of instantaneous elastic moduli, $\delta p$ is the increment of the Lagrangian multiplier that imposes the incompressibility constraint. The two dots operator $(:)$ denotes the double contraction of the indices
\[
(\mathcal{A}_0 : \tens{\Gamma})_{ij} = (A_0)_{ijhk}\Gamma_{kh},
\]
where the convention of summation over repeated indices is used.
The components of the tensor $\mathcal{A}_0$ for a neo-Hookean material are given by 
\[
(A_0)_{ijhk} = \mu \delta_{ik}(B_\text{e})_{ih} 
\]
where $\tens{B}_\text{e} = \tens{F}_\text{e} \tens{F}_\text{e}^T$ and $\delta_{ik}$ is the Kronecker delta. The incremental equilibrium equation and the linearised form of the incompressibility constraint read respectively
\begin{equation}
\left\{
\begin{aligned}
\label{eq:deltaP}
&\Diver\delta\tens{P} = \vect{0} &&\hbox{in } \Omega_{0\text{L}}, \,\Omega_{0\text{C}}, \\
&\tr \, \tens{\Gamma} = 0 &&\hbox{in } \Omega_{0\text{L}},\, \Omega_{0\text{C}}.
\end{aligned}
\right.
\end{equation}
The linearised form of the kinematic constraint \eqref{eq:BCcenter} and of the boundary condition \eqref{eq:BC_capi_T} complement the incremental equations 
\begin{gather}
\label{eq:BC_incrementedu}
\delta\vect{u}(\vect{0}) = \vect{0},\\
\label{eq:BC_incrementedP}
\delta \tens{P}\, \vect{e}_r = \gamma\delta\mathcal{K} \,\vect{e}_r - \gamma\mathcal{K} \tens{\Gamma}^T\, \vect{e}_r,
\end{gather}
where $\delta \mathcal{K}$ is the increment of the curvature. Finally, we enforce the continuity of the incremental displacement of the stress at the interface
\begin{equation}
\label{eq:cont_incr}
\left\{
\begin{aligned}
&\lim_{r\rightarrow R_\text{i}^-}\delta\vect{u} = \lim_{r\rightarrow R_\text{i}^+}\delta\vect{u},\\
&\lim_{r\rightarrow R_\text{i}^-}\delta\tens{P}^T\vect{e}_r = \lim_{r\rightarrow R_\text{i}^+}\delta\tens{P}^T\vect{e}_r.
\end{aligned}
\right.
\end{equation}

In the following, we rewrite the incremental problem given by the Eqs. \eqref{eq:deltaP}-\eqref{eq:BC_incrementedP} into a more convenient form using the \cite{Stroh_1962} formulation.

\subsection{Stroh formulation}

We rewrite the incremental problem in non-dimensional form adopting as the characteristic length scale and shear modulus $R_\text{o}$ and $\mu$ respectively. The behaviour of the problem is governed by the non-dimensional parameters
\begin{equation}
\label{eq:par_adim}
\alpha_\gamma = \frac{\ell_{\text c}}{R_{\text o}}= \frac{\gamma}{\mu R_\text{o}}, \quad \alpha_R = \frac{R_{\text i}}{R_{\text o}},
\end{equation}
in addition to the growth parameter $g$.\\
For the sake of brevity, we introduce the multi-index ${\text W} = \{{\text L},{\text C}\}$. The quantities with subscript ${\text L}$ are computed in the lumen, while the ones in the cortex have the subscript ${\text C}$.

We denote with $u_{\text W}$ and $v_{\text W}$ the components of $\delta \vect{u}_{\text W}$ while $\delta P^{\text W}_{rr}$ and $\delta P^{\text W}_{r\theta}$ are the components of the incremental stress along the radial direction $\vect{e}_r$. We can reduce the system of partial differential equations \eqref{eq:deltaP} into a system of ordinary differential equations by assuming the following ansatz for the incremental displacement, pressure and stress:
\begin{align}
\label{eq:ul}
u_{\text W}(r,\theta)&=U_{\text W}(r) \cos(m\theta),\\
\label{eq:vl}
v_{\text W}(r,\theta)&=V_{\text W}(r) \sin(m\theta),\\
\label{eq:deltaprr}
\delta P^{\text W}_{rr}(r,\theta)&= s_{rr}^{\text W}(r) \cos(m\theta),\\
\label{eq:deltaprteta}
\delta P^{\text W}_{r\theta}(r,\theta)& = s_{r\theta}^{\text W}(r) \sin(m\theta),\\
\label{eq:deltapres}
\delta p^{\text W}(r,\theta)& = Q^{\text W}(r) \cos(m\theta),
\end{align}
where $m\in \{n\in\mathbb{N}\;|\;n\geq 2\}$ is the circumferential wavenumber.
By substituting Eqs. \eqref{eq:ul}-\eqref{eq:deltaprr} and Eq.~\eqref{eq:deltapres} into Eq. \eqref{eq:deltaP}, we obtain the following expression for $Q^{\text L}$ and $Q^{\text C}$
$$
\begin{aligned}
Q^{\text L}(r) &= - s_{rr}^{\text L}(r)+p_{\text L}U'_{\text L}(r),\\
Q^\Cped (r)&=U'_{\Cped}(r) \left(\frac{\alpha_R^2
   \left(g^2-1\right)}{r^2}+p_{\text C}+1\right)-s_{r\theta}^{C}(r),
\end{aligned}
$$
where $p_\text{L}$, $p_\text{C}$ are defined in Eq.~\eqref{eq:ri<r<ro}.

By the choices \eqref{eq:ul}-\eqref{eq:deltaprteta} and using a well established procedure, the incremental problem can be rewritten in the \cite{Stroh_1962} form
\begin{equation}
\label{eq:etalumen}
\frac{d \vect{\eta}_{\text W}}{d r} = \frac{1}{r} \tens{N}_{\text W} \vect{\eta}_{\text W} \quad \hbox{with } {\text W = \{ \text{L} ,\,\text{C}\}},
\end{equation}
where $\vect{\eta}_{\text W}$ is the \emph{displacement-traction vector} defined as
\[
\vect{\eta}_{\text W} = [\vect{U}_{\text W}, r\vect{\Sigma}_{\text W}] \quad \hbox{where} \quad \left\{
\begin{aligned}
&\vect{U}_{\text W} = [U_{\text W},V_{\text W}],\\
&\vect{\Sigma}_{\text W}= [s_{rr}^{\text W}, s_{r\theta}^{\text W}].
\end{aligned}
\right.
\]
The matrix $\tens{N}_{\text W} \in \mathbb{R}^{4\times4}$ is the \emph{Stroh matrix} and it has the following sub-block form
\[
\tens{N}_L=
\begin{bmatrix}
\tens{N}^{\text W}_1 &\tens{N}^{\text W}_2\\
\tens{N}^{\text W}_3 &\tens{N}^{\text W}_4
\end{bmatrix}.
\]
For the lumen ($r< r_\text{i}$), the sub-blocks read:
\begin{equation}
\label{eq:strohlumen}
\begin{aligned}
&\tens{N}^{\text L}_1=
\begin{bmatrix}
 -1 & -m \\
 mp_{\Lped} & p_\Lped
 \end{bmatrix},
 &&\tens{N}^{{\text L}}_3 = 
\begin{bmatrix}
 p_\text{L} \left(2-m^2 p_\text{L}\right)+m^2+2 & -m (p_\text{L}-3) (p_\text{L}+1) \\
 -m (p_\text{L}-3) (p_\text{L}+1) & (p_\text{L}+1) \left(2 m^2-p_\text{L}+1\right)
\end{bmatrix}
\\
 &\tens{N}^{\text L}_2= 
 \begin{bmatrix}  
0  & 0\\
 0 & 1 \\\end{bmatrix},
  &&\tens{N}^{\text L}_4 = 
\begin{bmatrix} 1 &-m p_\text{L}\\ m &- p_\text{L} \end{bmatrix}
\end{aligned}
\end{equation}
In the cortex ($r_\text{i}<r<r_\text{o}$), the sub-blocks are given by
\begin{equation}
\label{eq:strohcortex}
\begin{gathered}
\tens{N}^\Cped_1=
\begin{bmatrix}
 -1 & -m \\
 mp_\Cped\beta_\Cped & p_\Cped\beta_\Cped
 \end{bmatrix},\qquad
 \tens{N}^\Cped_2= 
 \begin{bmatrix}  
 0  & 0\\
 0 & \beta_{\Cped} \\\end{bmatrix},\qquad
 \tens{N}^\Cped_4 = 
 \begin{bmatrix}  
  1  &-mp_\Cped\beta_\Cped\\
  m & -p_\Cped\beta_\Cped \end{bmatrix},\\
 \tens{N}^{\Cped}_3 = 
 \begin{bmatrix}  
 1/\beta_\Cped+(1+m^2)\beta_\Cped+p_\Cped(2-m^2 p_\Cped\beta_\Cped)& m\beta_\Cped\left(2+1/\beta_\Cped^2+2p_\Cped/\beta_\Cped-p_\Cped^2\right)\\
  m\beta_\Cped\left(2+1/\beta_\Cped^2+2p_\Cped/\beta_\Cped-p_\Cped^2\right) &  m^2/\beta_{\Cped}+(1+m^2)\beta_{\Cped}+2m^2p_\Cped-p_\Cped^2\beta_\Cped\end{bmatrix};
\end{gathered}
\end{equation}
where 
\[
\beta_\text{C} = \frac{r^2}{r^2+(g^2-1)\alpha_R^2}.
\]
We remark that the coefficient of the Stroh matrix are constant in the lumen (see Eq.~\eqref{eq:strohlumen}). This allows us to solve analytically the incremental problem for $r<r_\text{i}$.

\subsection{Incremental solution for the lumen}
We follow the procedure proposed in \citep{Dervaux_2011} and in  \citep{balbi2018mechanics}. Since Eq.~\eqref{eq:etalumen} with $\text{W} = \text{L}$ is a system of ODEs with constant coefficients, its solution can be rewritten in terms of eigenvalues and eigenvectors of $\tens{N}_{\text L}$.
The eigenvalues of $\tens{N}_{\text L}$ are $\lambda_1 = m-1$, $\lambda_2 = m+1$, $\lambda_3 = -m+1$ and $\lambda_4 = -m-1$. The general integral of Eq.~\eqref{eq:etalumen} is given by
\begin{equation}
\label{eq:sollumen}
\vect{\eta}_{\text L} = c_1 \vect{w}_1 r^{m-1} + c_2 \vect{w}_2 r^{m+1}+c_3 \vect{w}_3 r^{-m+1} + c_4 \vect{w}_2 r^{-m-1},
\end{equation}
where $\vect{w}_i$ are the eigenvectors of $\tens{N}_\text{L}$ associated with the eigenvalues $\lambda_i$, $i = 1,\dots,\,4$. Since the incremental solution must satisfy the kinematic constraint Eq.~\eqref{eq:BC_incrementedu}, we immediately get that $c_3 = c_4 = 0$ while
\begin{equation}
\label{eq:w_i}
\begin{aligned}
\vect{w}_1 &= \left[-1,1,-(m-1) (1+p_\text{L}),(m-1) (1+p_\text{L})\right],\\
\vect{w}_2 &=\left[-m,m+2,-(m+1) (m (1+p_\text{L})-4),(m+1) ((m+2)+(m-2) p_\text{L})\right].
\end{aligned}
\end{equation}
The two constants $c_1$ and $c_2$ will be fixed by imposing the continuity of the displacement and of the stress at $r = r_{\text i}$ (i.e. by enforcing Eq.~\eqref{eq:cont_incr}).

\subsection{Numerical procedure for the solution in the cortex}
The incremental problem in the cortex cannot be solved analytically since the coefficient of the Stroh matrix $\tens{N}_\text{C}$ given by Eq.~\eqref{eq:strohcortex} are not constant.
To overcome this difficulty, we implement a numerical code based on the impedance matrix method \citep{biryukov1985impedance, Biryukov_1995}.

We first introduce the \emph{conditional impedance matrix}, defined as
\begin{equation}
\label{eq:impedancematrix}
r \vect{\Sigma}_\Cped(r) = \tens{Z}_{\Cped}(r,r_{\text o})\vect{U}_{\Cped}(r).
\end{equation}
Such a matrix is called conditional since its expression depends on an auxiliary condition at $r = r_\text{o}$, in this case the boundary condition Eq.~\eqref{eq:BC_incrementedP}. In the following paragraphs we expose a procedure to construct the matrix $\tens{Z}_\text{C}(r,\,r_\text{o})$.
The incremental boundary condition Eq.~\eqref{eq:BC_incrementedP} reads
\begin{equation}
\label{eq:BC_incrementedP2}
\delta\tens{P}^T \vect{e}_{r} = \alpha_\gamma \delta\mathcal{K}\, \vect{e}_{r} +\frac{\alpha_\gamma}{r_{\text o}}\tens{\Gamma}^T\,\vect{e}_r,\qquad \text{at }r = r_\text{o}.
\end{equation}
where the incremental curvature $\delta\mathcal{K}$ is given by (see the Appendix~\ref{app:A} for the details of the computation):
\[
\delta\mathcal{K} = \frac{1}{r_\text{o}^2}\left(\dfrac{\partial u}{\partial \theta}+\dfrac{\partial^2 u}{\partial \theta^2}\right),\qquad \text{at }r = r_\text{o}.
\] 
From the ansatz of variable separation in Eqs.\eqref{eq:ul}-\eqref{eq:vl} and using the incremental form of the incompressibility constraint Eq.~\eqref{eq:deltaP}, the boundary condition Eq.~\eqref{eq:BC_incrementedP2} is equivalent to
\begin{equation}
\label{eq:P_tens_sup}
\delta\tens{P}^T \vect{e}_{r}= -\frac{\alpha_\gamma}{r_{\text o}^2}\left[(m^2U_\Cped+mV_\Cped)\cos (m\theta),(mU_\Cped+V_\Cped)\sin (m\theta)\right]\qquad\text{at }r=r_\text{o}.
\end{equation}

We can now define the \emph{auxiliary impedance matrix} \citep{Norris_2010} as
\begin{equation}
\label{eq:zi}
\tens{Z}_{\text o} = -\frac{\alpha_\gamma}{r_{\text o}} \begin{bmatrix}
m^2 & m\\
m &1
\end{bmatrix},
\end{equation}
so that the boundary condition Eq.~\eqref{eq:P_tens_sup} is equivalent to the equation
\begin{equation}
\label{eq:auxiliary_condition}
r_\text{o}\vect{\Sigma}_\text{C}(r_\text{o}) = \tens{Z}_\text{o}\vect{U}(r_\text{o}).
\end{equation}
We introduce the matricant
\[
\tens{M}^{\Cped}(r,r_{\text o}) = \begin{bmatrix}
\tens{M}^{\Cped}_1(r,r_{\text o}) &\tens{M}^{\Cped}_2(r,r_{\text o})\\
\tens{M}^{\Cped}_3(r,r_{\text o})&\tens{M}^{\Cped}_4(r,r_{\text o})
\end{bmatrix}, \qquad \tens{M}^{\Cped}(r,r_{\text o}) \in \R^{4\times4}
\]
called \emph{conditional matrix}, defined as the solution of the problem
\begin{equation}
\label{eq:matricant}
\left\{
\begin{aligned}
&\frac{d\tens{M}^{\Cped}}{dr} = \frac{1}{r}\tens{N}_\Cped\tens{M}^{\Cped}(r,r_{\text o})\\
&\tens{M}^\text{C}(r_{\text o},r_{\text o}) = \tens{I}.
\end{aligned}
\right.
\end{equation}
From Eq.~\eqref{eq:matricant}, the Stroh form of the incremental problem given by Eq.~\eqref{eq:etalumen} and Eq.~\eqref{eq:auxiliary_condition}, we get
\[
\left\{
\begin{aligned}
&\vect{U}_\Cped(r) = (\tens{M}_1^\text{C}(r,r_{\text o}) + r_\text{o}\tens{M}_2^\text{C}(r,r_{\text o})\tens{Z}_\text{o})\vect{U}_\Cped(r_{\text o}),\\
&r\vect{\Sigma}_\Cped(r) = (\tens{M}_3^\text{C}(r,r_{\text o}) + r_\text{o}\tens{M}_4^\text{C}(r,r_{\text o})\tens{Z}_\text{o})\vect{U}_\Cped(r_{\text o}),
\end{aligned}
\right.
\]
so that
\begin{equation}
\label{eq:schifo}
r\vect{\Sigma}_\Cped(r) = (\tens{M}_3^\text{C}(r,r_{\text o}) + r_\text{o}\tens{M}_4^\text{C}(r,r_{\text o})\tens{Z}_\text{o})(\tens{M}_1^\text{C}(r,r_{\text o}) + r_\text{o}\tens{M}_2^\text{C}(r,r_{\text o})\tens{Z}_\text{o})^{-1}\vect{U}_\Cped(r).
\end{equation}
From Eq.~\eqref{eq:schifo}, the conditional impedance matrix is given by
\[
\tens{Z}_\text{C}(r,\,r_\text{o}) = (\tens{M}_3^\text{C}(r,r_{\text o}) + r_\text{o}\tens{M}_4^\text{C}(r,r_{\text o})\tens{Z}_\text{o})(\tens{M}_1^\text{C}(r,r_{\text o}) + r_\text{o}\tens{M}_2^\text{C}(r,r_{\text o})\tens{Z}_\text{o})^{-1}.
\]

From now on, we omit the dependence of $\tens{Z}_\text{C}$ wherever convenient for sake of simplicity. By using Eq.~\eqref{eq:impedancematrix}, we can rewrite the Stroh problem given by Eq.~\eqref{eq:etalumen} into a differential Riccati equation. Indeed, from Eq.~\eqref{eq:etalumen}, we get
\begin{gather}
\label{eq:riccatiU}
\frac{d \vect{U}_\Cped}{d r} = \frac{1}{r} \left(\tens{N}_1^\Cped\vect{U}_\Cped+\tens{N}_2^\Cped\tens{Z}_\Cped\vect{U}_\Cped\right),\\
\label{eq:riccatiZU}
\frac{d\tens{Z}_\Cped}{dr}\vect{U}_\Cped+\tens{Z}_\Cped\frac{d\vect{U}_\Cped}{dr} = \frac{1}{r} \left(\tens{N}_3^\Cped\vect{U}_\Cped+\tens{N}_4^\Cped\tens{Z}_\Cped\vect{U}_\Cped\right).
\end{gather}
Substituting Eq.~\eqref{eq:riccatiU} into Eq.~\eqref{eq:riccatiZU} we get the following differential Riccati equation
\begin{equation}
\label{eq:riccati}
\frac{d \tens{Z}_\Cped}{d r} = \frac{1}{r} \left(\tens{Z}_\Cped\tens{N}_1^\Cped-\tens{Z}_\Cped\tens{N}_2^\Cped\tens{Z}_\Cped+\tens{N}_3^{\Cped}+\tens{N}_4^{\Cped}\tens{Z}_\Cped\right).
\end{equation}
We integrate Eq.~\eqref{eq:riccati} from $r_{\text o}$ to $r_{\text i}$, using as initial condition the auxiliary impedance matrix defined in Eq.~\eqref{eq:zi}, i.e.
\[
\tens{Z}_\text{c}(r_\text{o},\,r_\text{o}) = \tens{Z}_\text{o}.
\]

To construct a bifurcation criterion, we follow \cite{balbi2018mechanics}: from the continuity of the displacement-traction vector $\vect{\eta}_\Cped(r_{\text i}) =\vect{\eta}_\Lped(r_{\text i})$ Eq.~\eqref{eq:cont_incr} we get
\[
r_{\text i}\vect{\Sigma}_\Lped(r_{\text i})= r_{\text i}\vect{\Sigma}_\Cped(r_{\text i}) = \tens{Z}_\Cped(r_{\text i},r_{\text o})\vect{U}_\Cped(r_{\text i}) =  \tens{Z}_\Cped(r_{\text i},r_{\text o})\vect{U}_\Lped(r_{\text i}),
\]
so that non-null solutions of the incremental problem exist if and only if
\begin{equation}
\label{eq:stopcondition}
\det \left[\tens{A}-\tens{Z}_\Cped(r_{\text i},r_{\text o})\tens{B}\right]=0,
\end{equation}
where 
\[
A_{ij} = (w_i)_{j+2} \qquad
B_{ij} = (w_i)_{j},\qquad i,j=1,2,
\]
with $\vect{w}_i$ defined in Eq.~\eqref{eq:w_i}.
For a fixed value of the control parameter $g$ we integrate the Riccati Eq.~\eqref{eq:riccati} from $r= r_{\text o}$ up to $r = r_{\text i}$ making use of the software \textsc{Mathematica} 11.3 (Wolfram Research, Champaign, IL, USA). We iteratively increase the control parameter $g$ until the bifurcation criterion Eq.~\eqref{eq:stopcondition} is satisfied.

\subsection{Discussion of the results}
\label{subsec:result_linear}

First, we need to identify an interval of interest for the adimensional parameter $\alpha_\gamma$. We have estimated the surface tension of cellular aggregates in Section~\ref{sec:surf_tens_spheroid}. From the stress profiles reported by \cite{Lee_2019}, we obtain a surface tension of the order of $10^{-1} \mathrm{N/m}$. According to \cite{Karzbrun_2018}, the shear modulus of the wild-type brain organoid is $\mu \simeq 900\,\mathrm{Pa}$ (Young modulus $E \simeq 2.7\,\mathrm{kPa}$) while $\mu \simeq 333\,\mathrm{Pa}$ (Young modulus $E \simeq 1\,\mathrm{kPa}$) for the unhealthy ones, afflicted by lissencephaly (i.e. when the mutation LIS1~$+/-$ is present). 
Provided that the typical radius of the organoid is about $R_\text{o}=400\,\mu\mathrm{m}$ \citep{Karzbrun_2018}, $\alpha_\gamma$ ranges between $0.25$ for the wild-type organoids and $0.75$ for the ones affected by lissencephaly.

Let us now discuss the results of the linear stability analysis.
For fixed values of the adimensional parameters $\alpha_R$ and $\alpha_\gamma$, we denote by $g_m$ the first value of $g$ such that the bifurcation criterion \eqref{eq:stopcondition} is satisfied for the wavenumber $m$. We define the critical threshold $g_\text{cr}$ as the minimum $g_m$ for $m\geq 2$ and the critical mode $m_\text{cr}$ as the wavenumber corresponding to $g_\text{cr}$.

\begin{figure}[b!]
	\begin{subfigure}{.5\linewidth}
		\centering
		\includegraphics[width=1\textwidth]{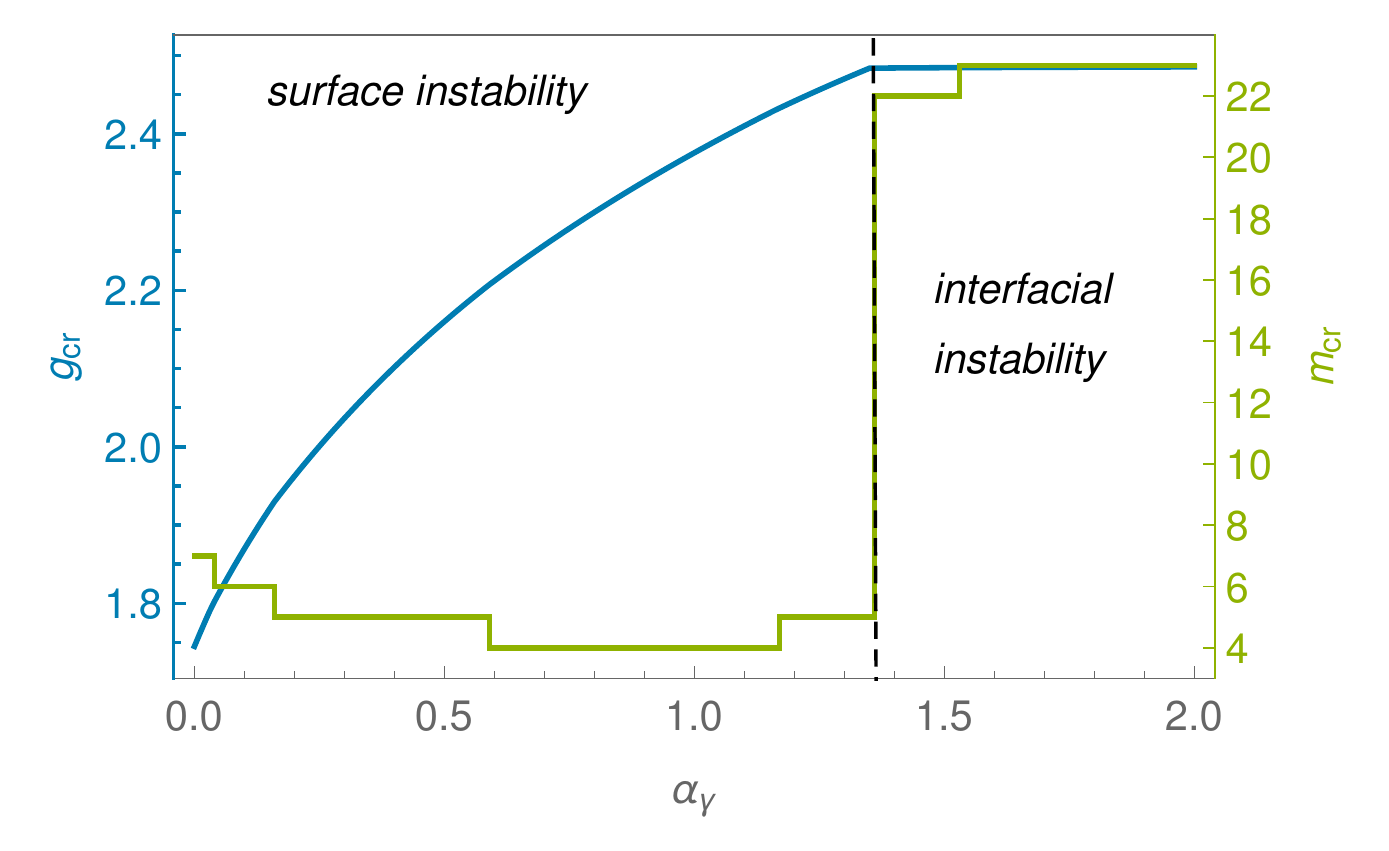}
		\caption{$\alpha_R = 0.9$}
		\label{fig:alphaR09}
	\end{subfigure}%
	\begin{subfigure}{.5\linewidth}
		\centering
		\includegraphics[width=1\textwidth]{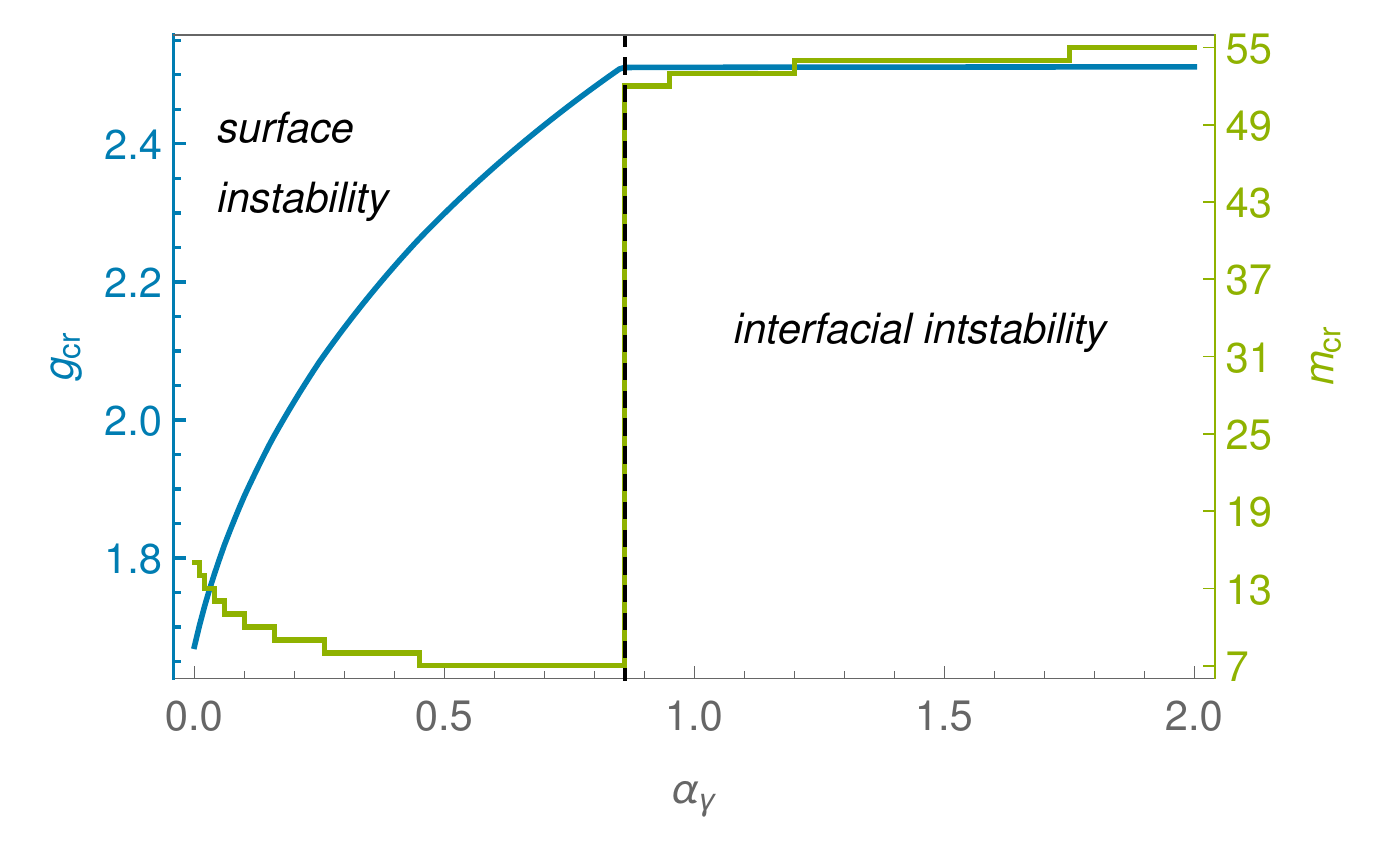}
		\caption{$\alpha_R = 0.95$}
		\label{fig:alphaR095}
	\end{subfigure}
	\caption{Plot of the marginal stability threshold $g_\text{cr}$ and of the critical mode $m_\text{cr}$ versus $\alpha_\gamma$ for (a) $\alpha_R=0.9$ and (b) $\alpha_R=0.95$.}
	\label{fig:fixalphaR}
\end{figure}

In Fig.~\ref{fig:fixalphaR}, we plot the critical values $g_{\text cr}$ and $m_{\text cr}$ versus $\alpha_{ \gamma}$ for two different values of $\alpha_R$, i.e. in the first one $\alpha_R = 0.9$, while in the other one $\alpha_R= 0.95$.
\begin{figure}[t!]
\centering
\includegraphics[width=0.5\textwidth]{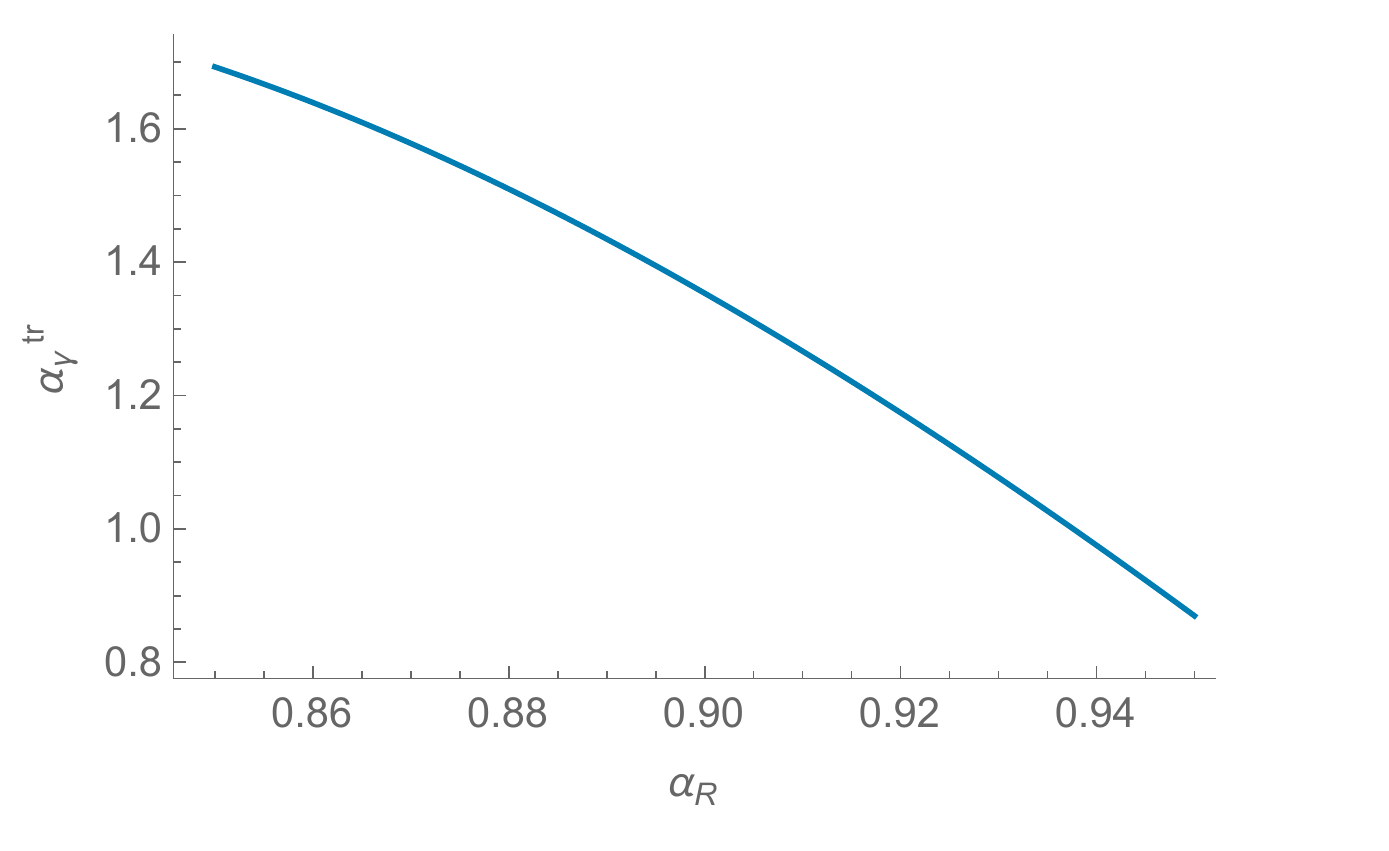}
\caption{Plot of the critical value $\alpha_\gamma^{\rm tr}$ at which the instability moves from the external surface to the interface versus $\alpha_R$.}
\label{fig:agcr_vs_ar}
\end{figure}
We observe that, for relatively small values of $\alpha_\gamma$, the marginal stability threshold $g_{\text cr}$ increases monotonously as $\alpha_\gamma$ increases, while the critical wavenumber $m_\text{cr}$ decreases. There is a change in the behaviour of the instability when the parameter $\alpha_\gamma$ is sufficiently large: the critical wavenumber $m_\text{cr}$ increases suddenly and the marginal stability threshold $g_{cr}$ remains nearly constant about $g_\text{cr}\simeq 2.5$.\\
The threshold at which this transition occurs strongly depends on $\alpha_R$, as shown in Fig.~\ref{fig:fixalphaR}. Indeed, when $\alpha_R= 0.9$, $g_{\text cr}$ increases from $1.745$ to $2.481$ with $\alpha_\gamma \in (0,1.34)$, while when $\alpha_R = 0.95$, $g_{\text cr}$ increases in approximately the same range as for $\alpha_R = 0.9$, i.e. $(1.671, 2.510)$, but $\alpha_\gamma$ varies in a smaller interval, i.e. $\alpha_\gamma \in (0,0.86)$. 

To study the morphology of the critical mode, we have integrated Eq.~\eqref{eq:riccatiU} to compute the incremental displacement field, as described in~\citep{Destrade_2009}. 
We depict in Tab.~\ref{tab:parametricplot} a morphological diagram where we show the solution of the incremental problem for different values of $\alpha_\gamma$ and $\alpha_R$. For small values of $\alpha_\gamma$, we observe that the instability mainly releases elastic energy at the free boundary, displaying a wrinkling pattern: as we increase $\alpha_\gamma$ the wavenumber decreases and the critical mode displays a more rounded boundary. Furthermore, for larger values of $\alpha_\gamma$ there is a drastic change in the features of the instability: the morphological transition localizes at the interface between the cortex and the lumen with a high critical wavenumber. Let us call $\alpha_\gamma^{\rm tr}$ the smallest value at which, fixing $\alpha_R$, the instability localizes at the interface between the cortex and the lumen. We plot in Figure~\ref{fig:agcr_vs_ar} $\alpha_\gamma^{\rm tr}$ versus $\alpha_R$. Interestingly, the transition from a surface to an interfacial instability can take place only if surface tension is relatively large. It is to be remarked that, whenever a surface instability takes place, the stretch induced by growth is quite large at the boundary between cortex and lumen. It is possible that, instead of wrinkles with short wavelength, interfacial creases appear. However, the nucleation of sharp creases is the result of a nonlinear instability, not detectable by a linear analysis \citep{Ciarletta_2019}.

\begin{table}[t!]
	\centering 
	\begin{tabular}{|c|c|c|c|c|}
		\hline
		& $\alpha_{\gamma} = 0$ &$\alpha_{\gamma} = 0.5$ &$\alpha_{\gamma} = 1$ &$\alpha_{\gamma} = 2$  \\ \hline
		
		$\alpha_R = 0.9$
		&\includegraphics[height=0.18\textwidth]{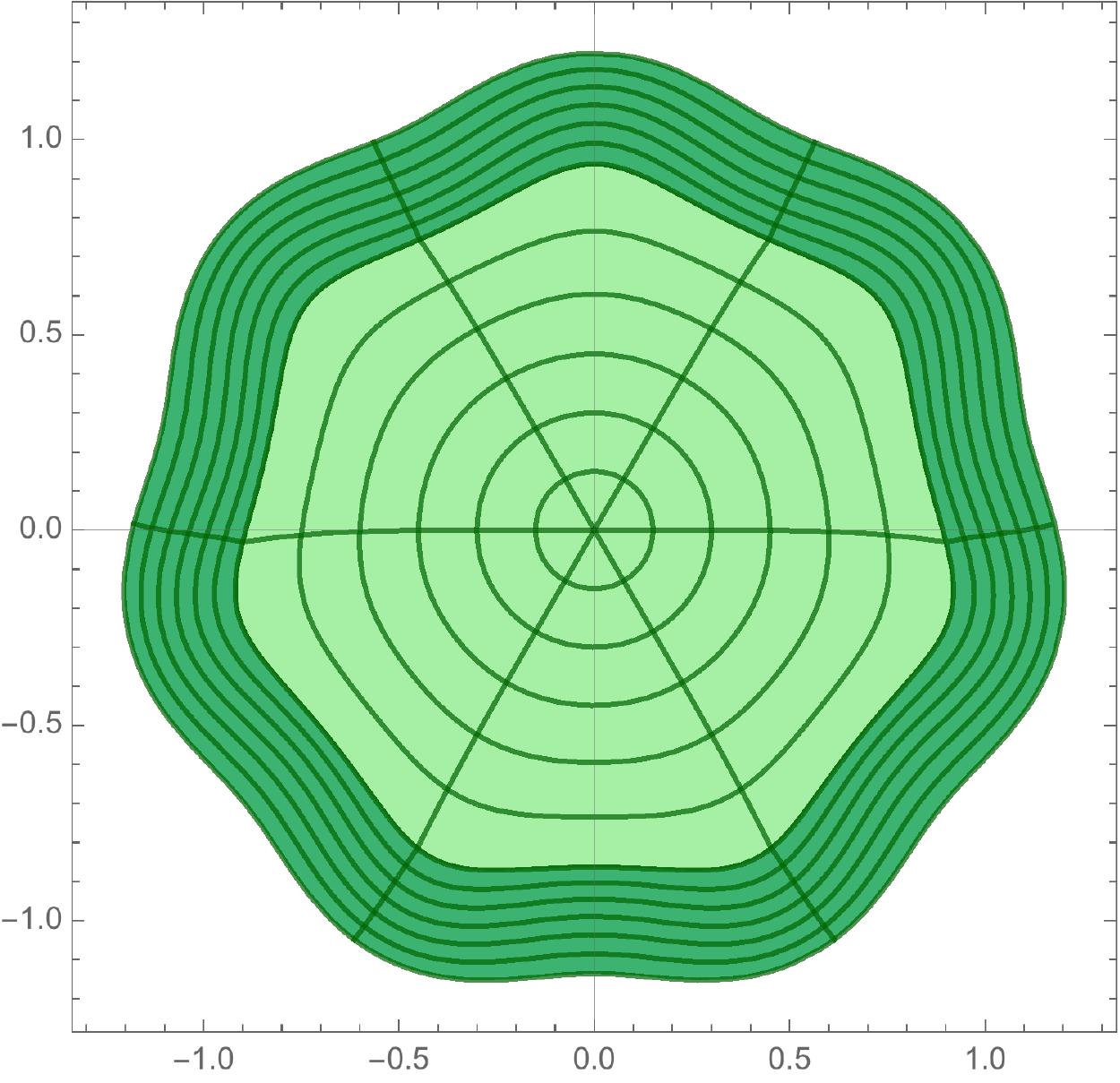}
		&\includegraphics[height=0.18\textwidth]{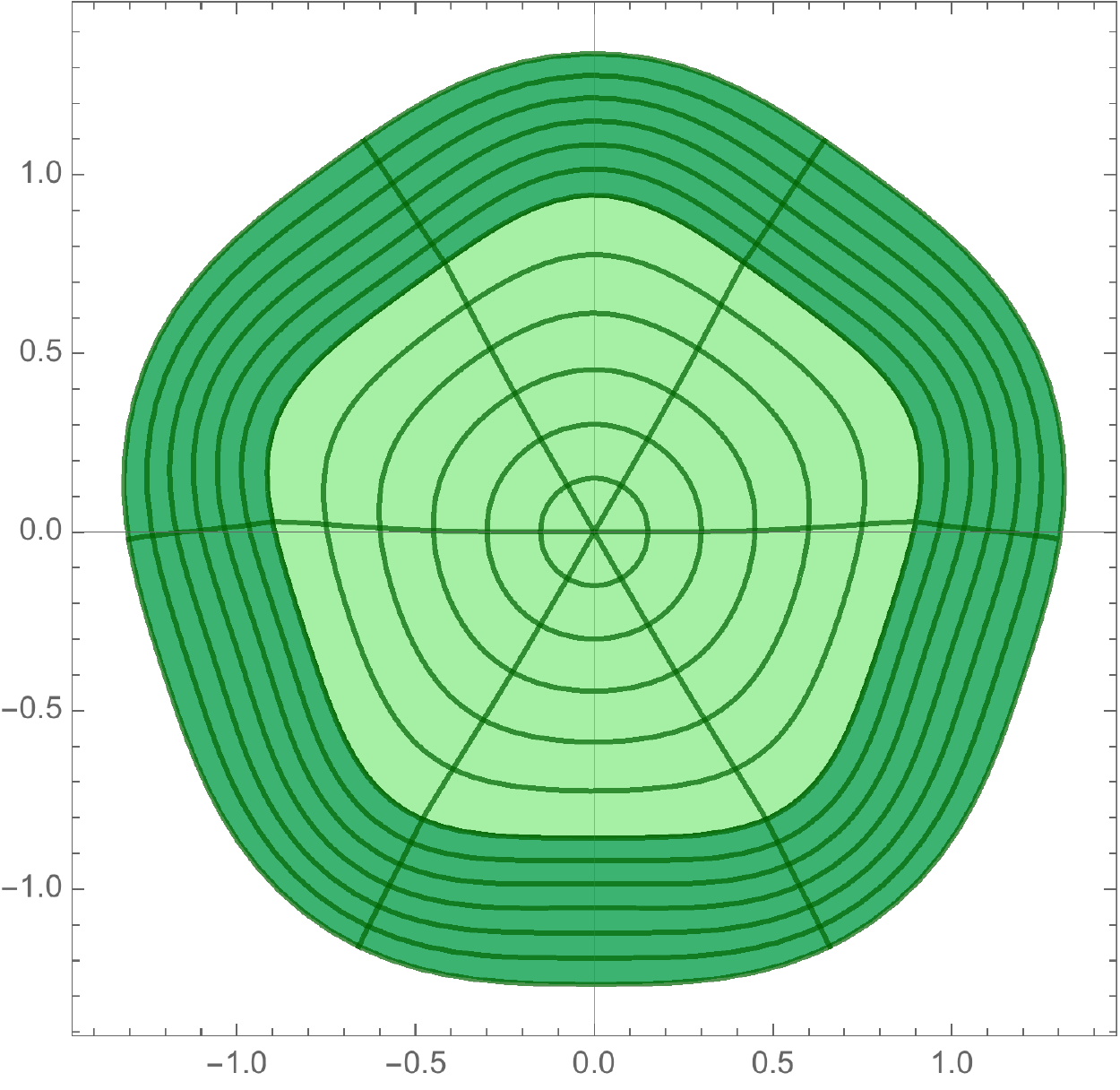}
		&\includegraphics[height=0.18\textwidth]{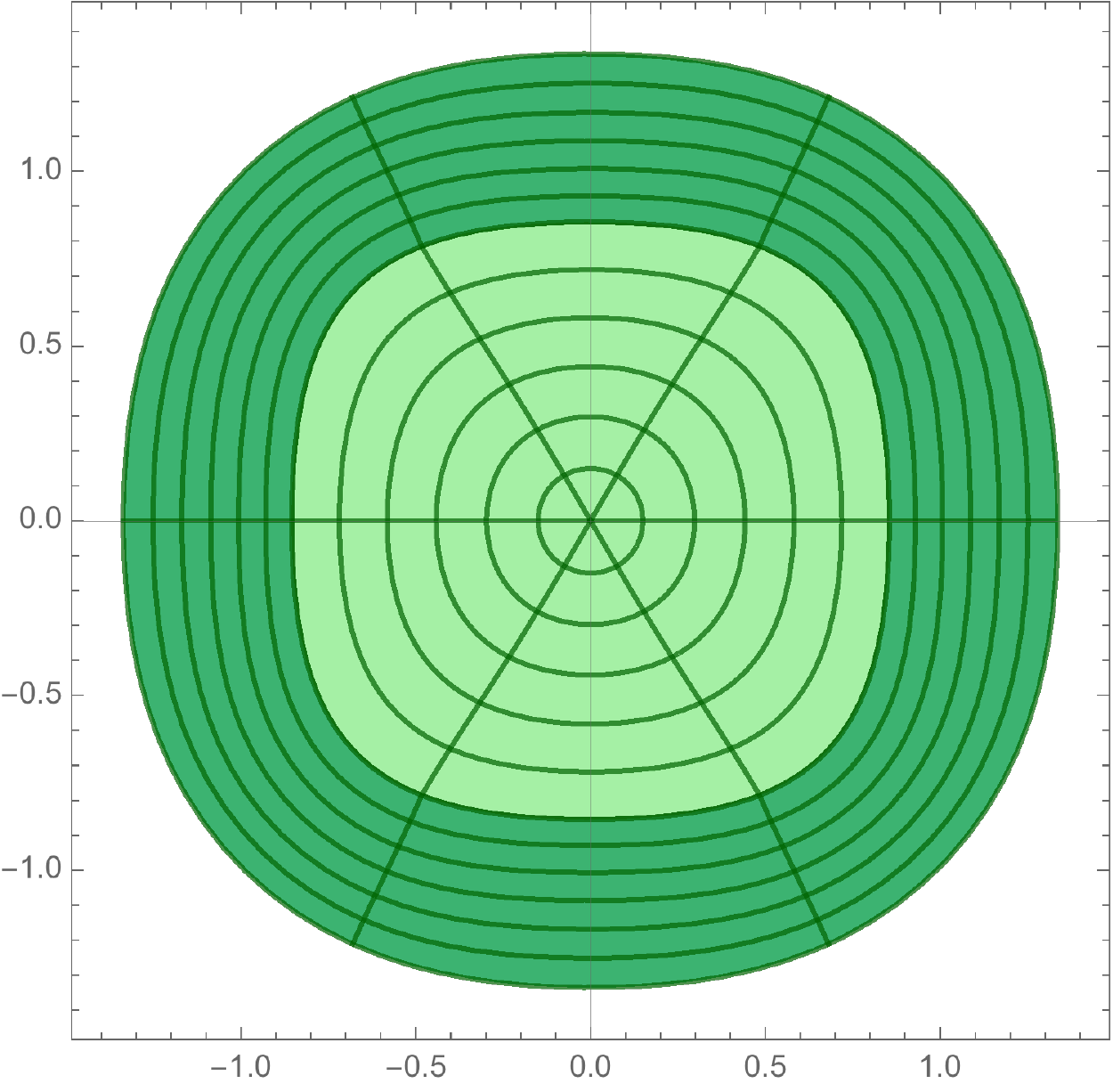}
		&\includegraphics[height=0.18\textwidth]{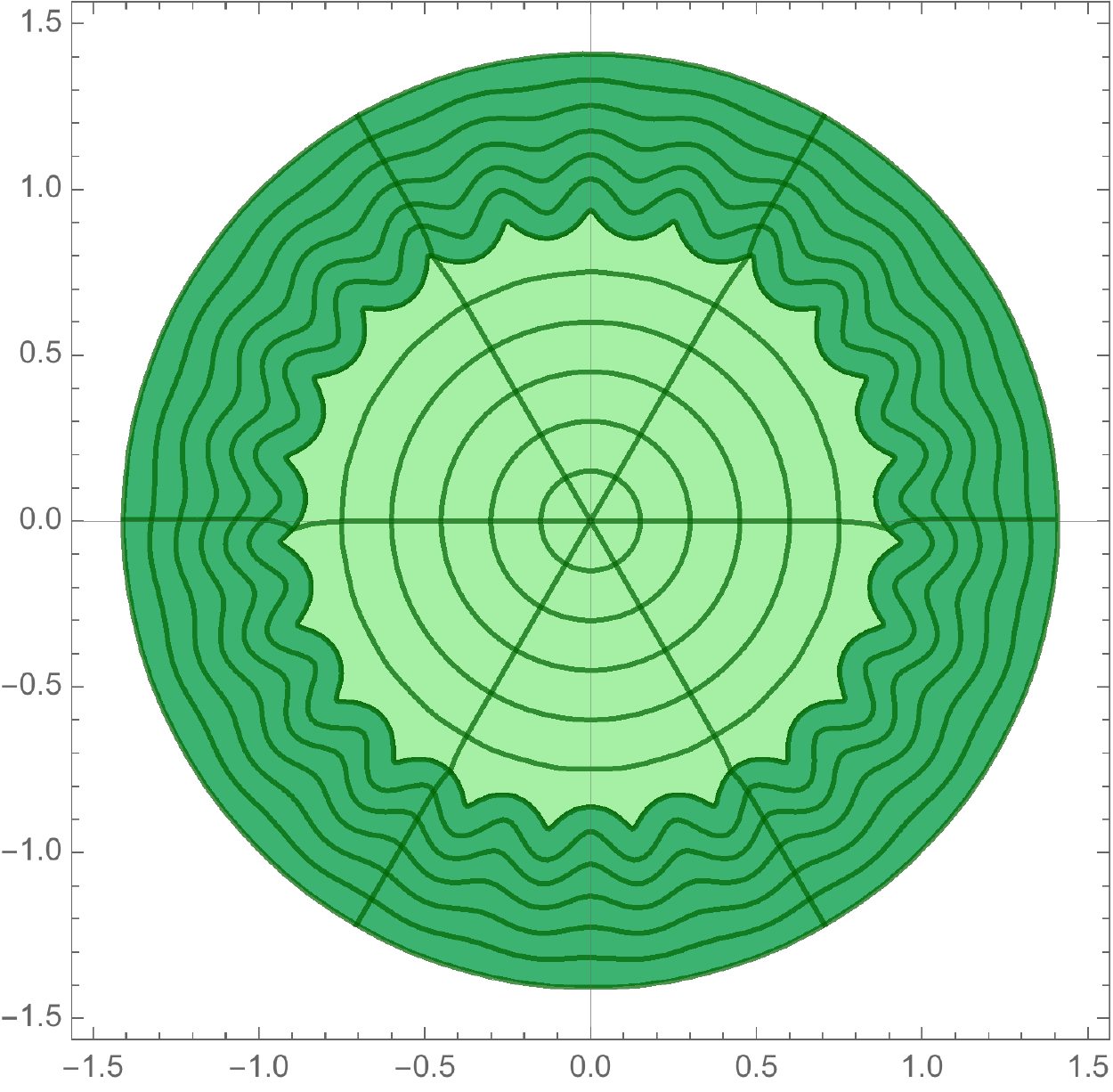}
		\\ \hline
		$\alpha_R = 0.95$
		&\includegraphics[height=0.18\textwidth]{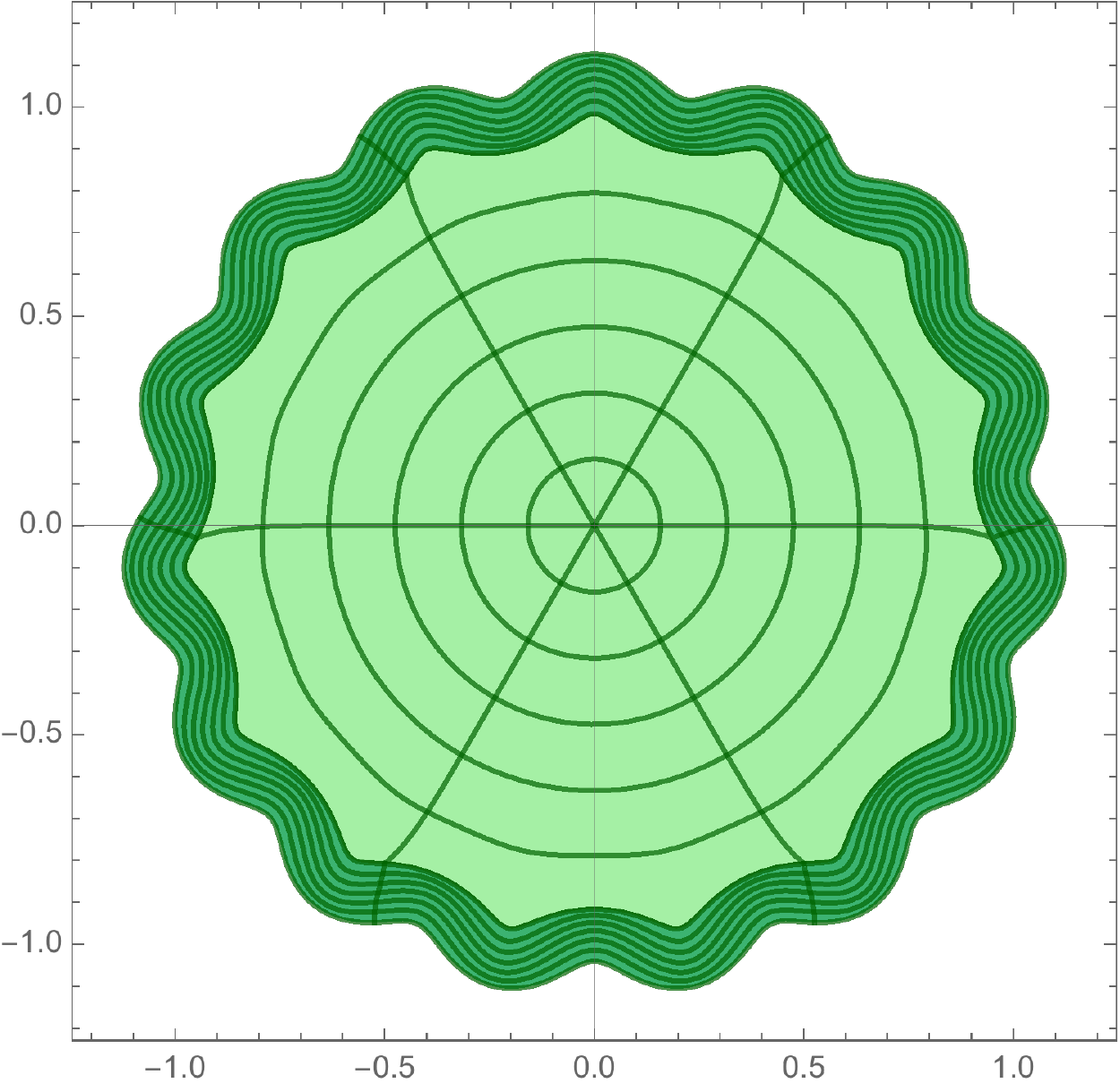}
		&\includegraphics[height=0.18\textwidth]{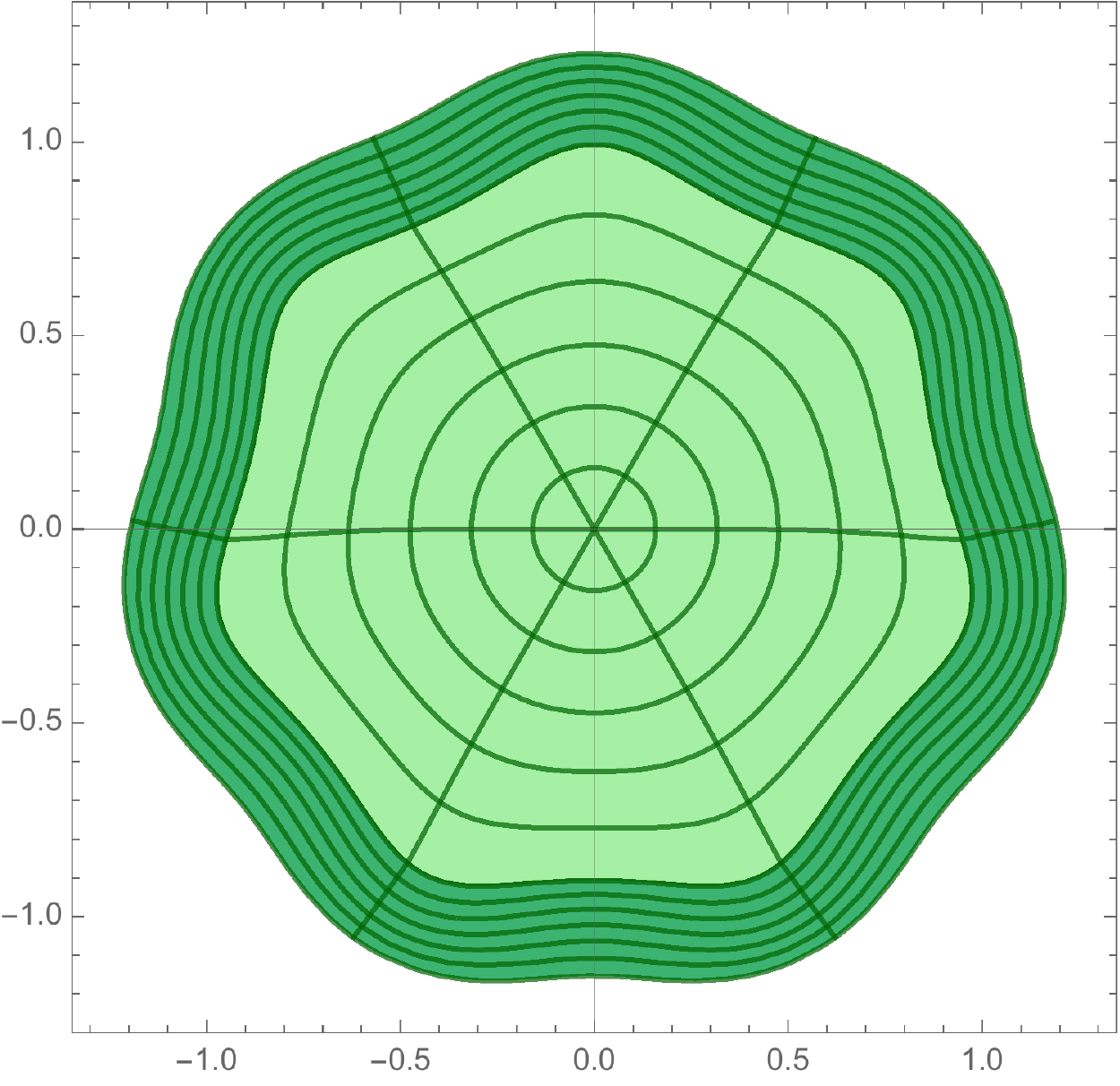}
		&\includegraphics[height=0.18\textwidth]{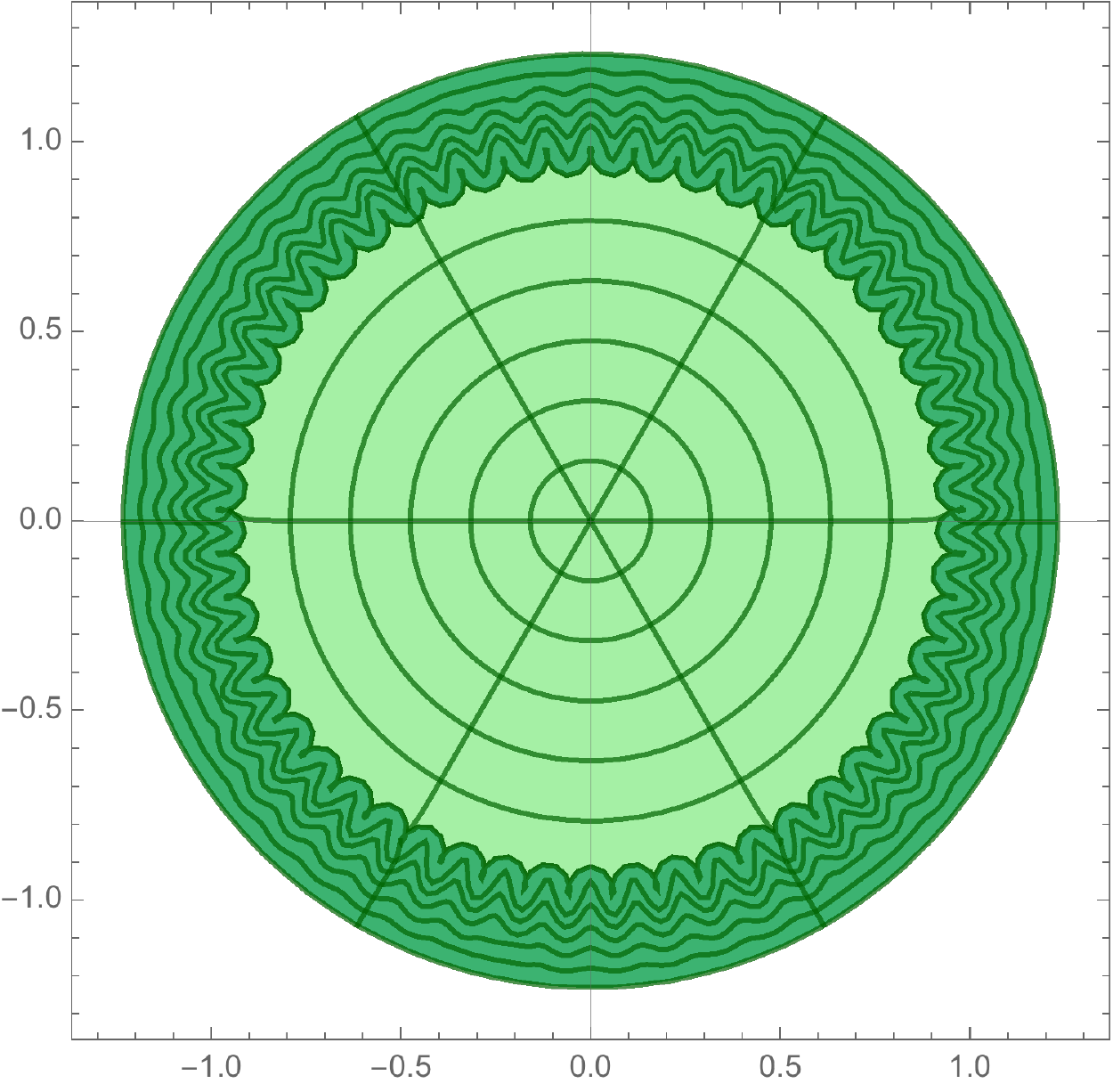}
		&\includegraphics[height=0.18\textwidth]{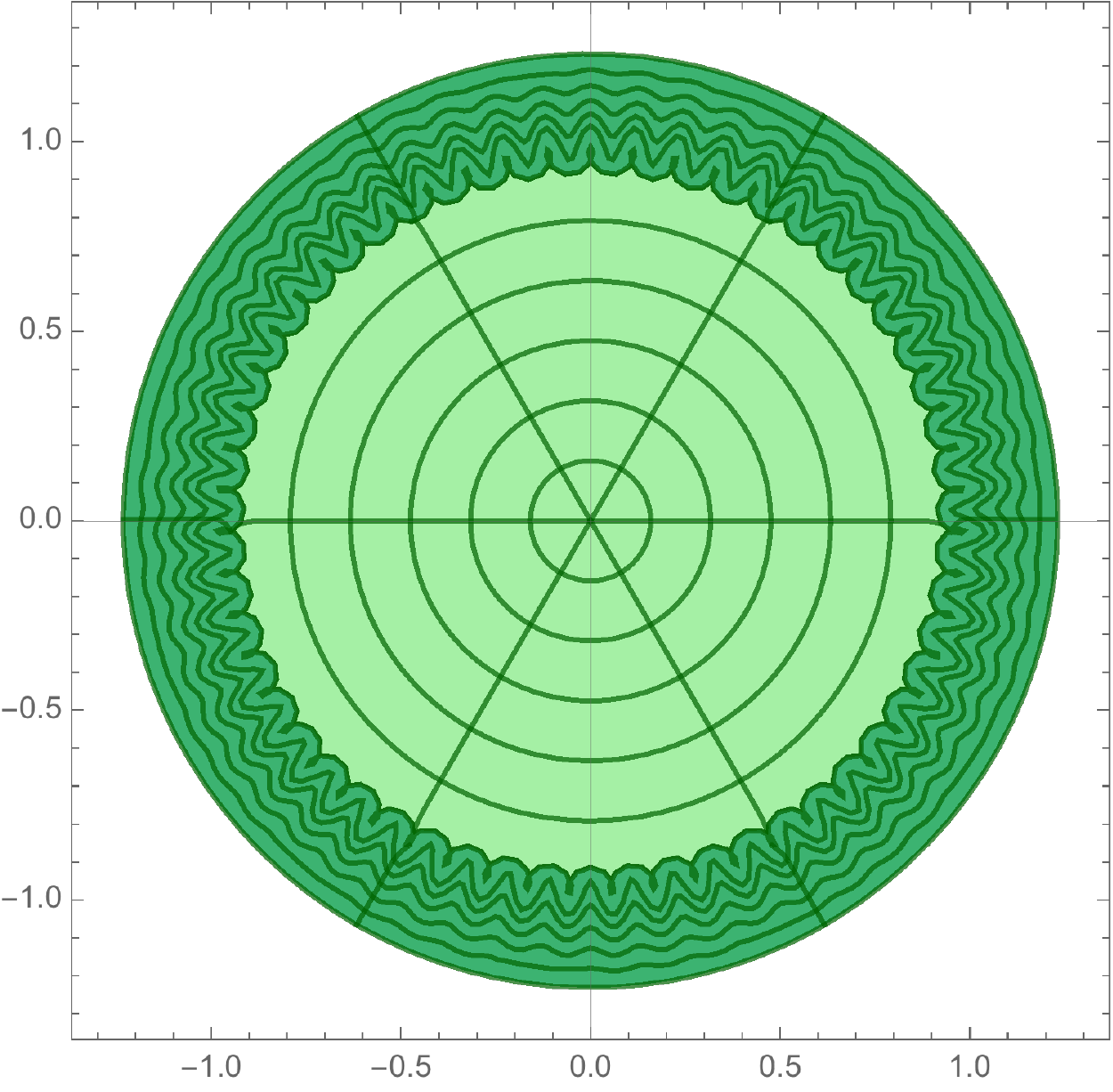}
		\\ \hline
	\end{tabular}
	\caption{Solutions of the linearised incremental problem at different $\alpha_R$ and different $\alpha_\gamma$. The amplitude of the incremental radial displacement has been set equal to $0.05 R_\text{o}$ for the sake of graphical clarity.
	}\label{tab:parametricplot}
\end{table}
\begin{figure}[h!]
	\begin{subfigure}{.5\linewidth}
		\centering
		\includegraphics[width=1\textwidth]{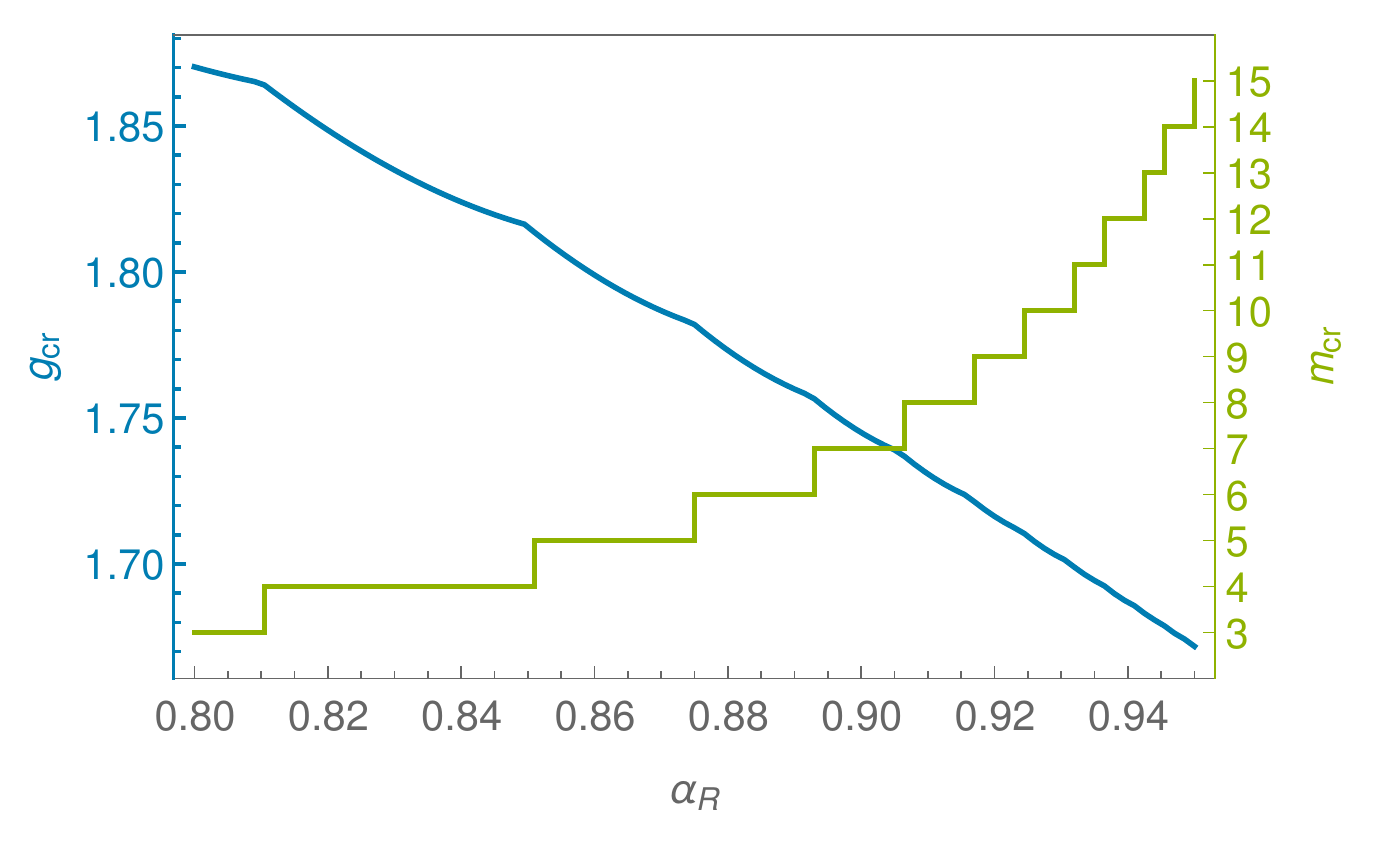}
		\caption{$\alpha_\gamma = 0$}
		\label{fig:alphagamma0}
	\end{subfigure}%
	\begin{subfigure}{.5\linewidth}
		\centering
		\includegraphics[width=1\textwidth]{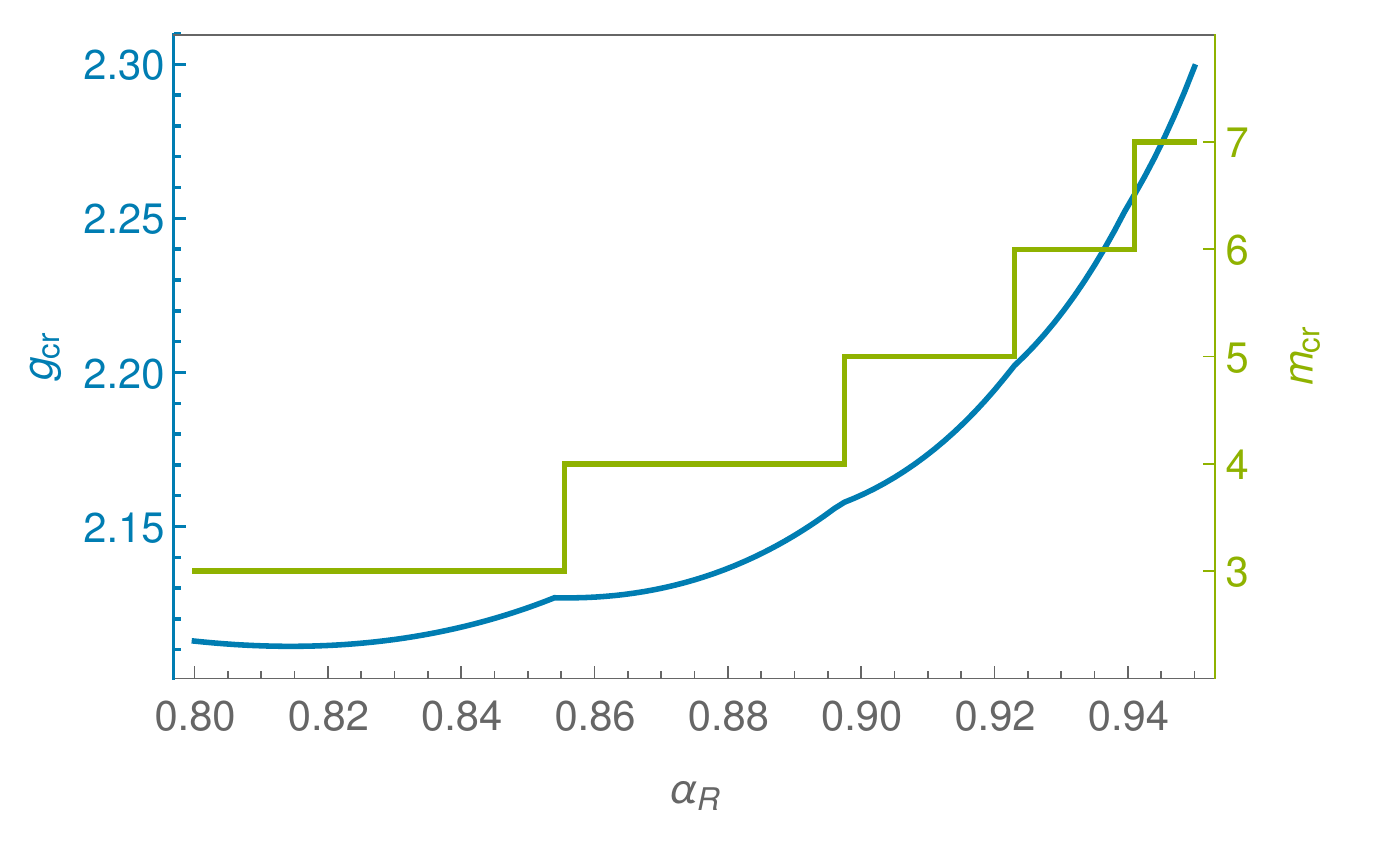}
		\caption{$\alpha_\gamma = 0.5$}
		\label{fig:alphagamma05}
	\end{subfigure}\\
	\begin{subfigure}{.5\linewidth}
		\centering
		\includegraphics[width=1\textwidth]{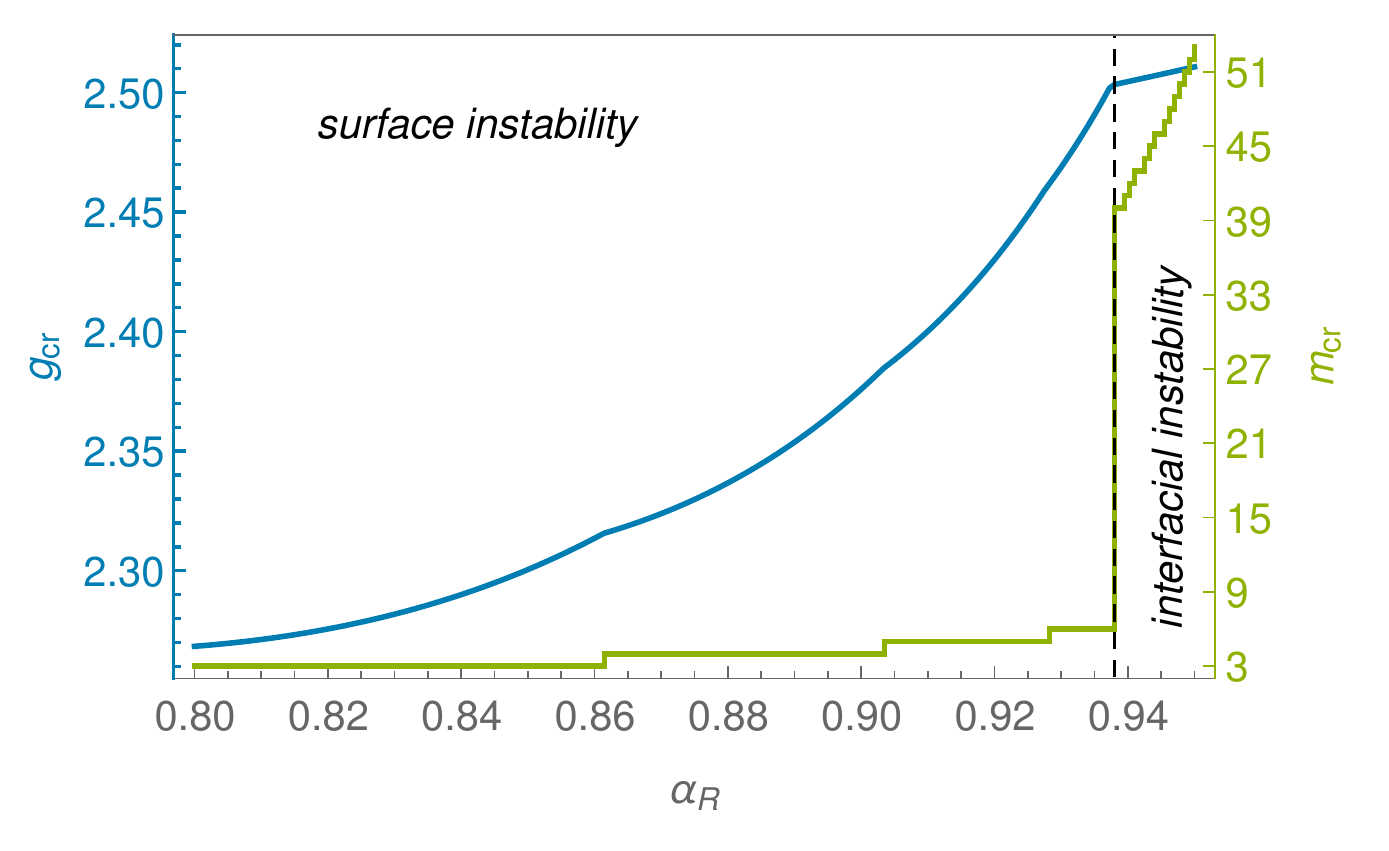}
		\caption{$\alpha_\gamma = 1$}
		\label{fig:alphagamma1}
	\end{subfigure}%
	\begin{subfigure}{.5\linewidth}
		\centering
		\includegraphics[width=1\textwidth]{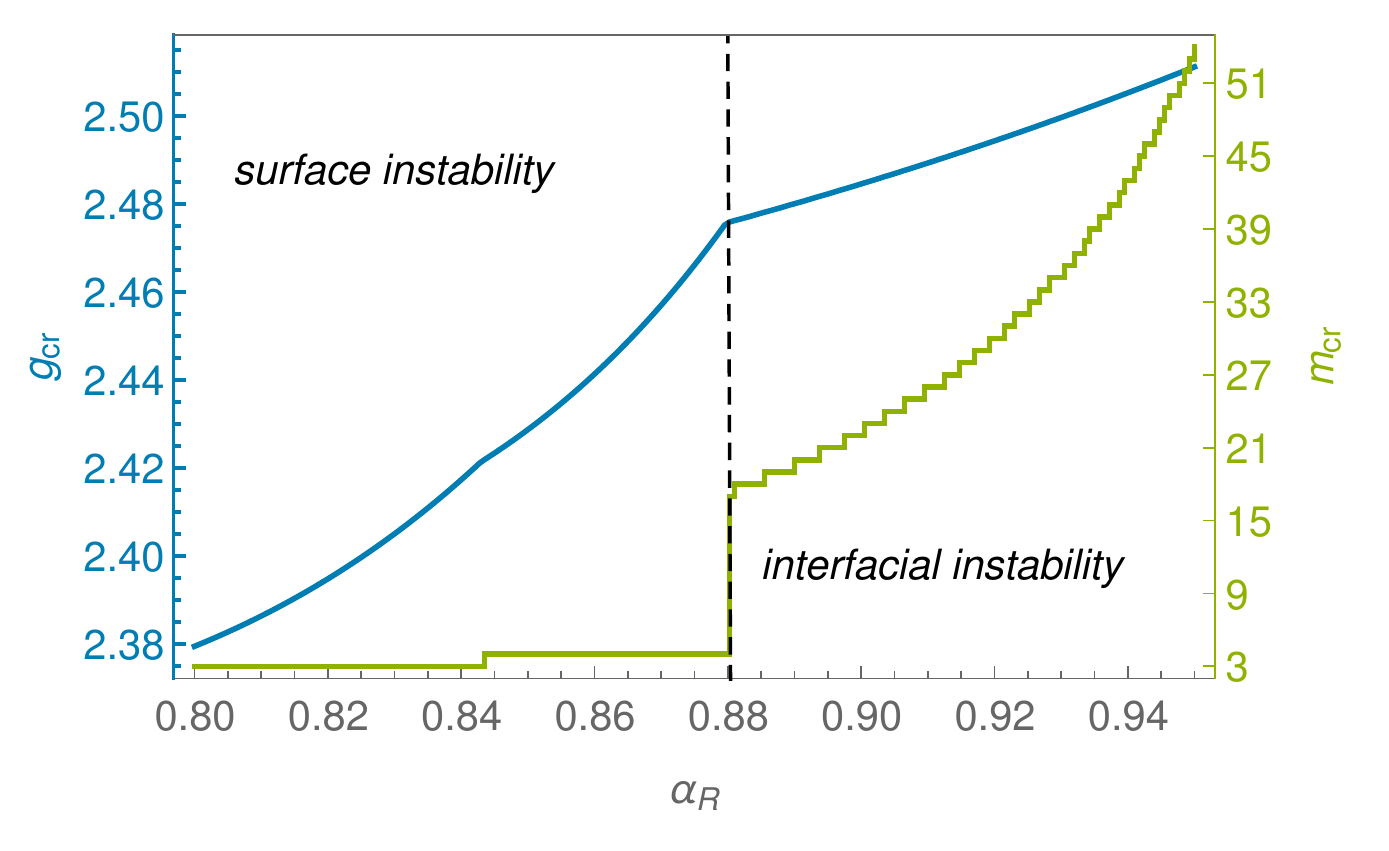}
		\caption{$\alpha_\gamma = 1.5$}
		\label{fig:alphagamma15}
	\end{subfigure}
	\caption{Plot of the marginal stability threshold $g_\text{cr}$ and of the critical mode $m_\text{cr}$ versus $\alpha_R$ for (a) $\alpha_\gamma=0$, (b) $\alpha_\gamma=0.5$, (c) $\alpha_\gamma=1$, (d) $\alpha_\gamma=1.5$.}
	\label{fig:fixalphagamma}
\end{figure}

We also investigate the influence of $\alpha_R$ on the instability by fixing $\alpha_\gamma$ (see Fig.~\ref{fig:fixalphagamma}). In the absence of surface tension (i.e. $\alpha_\gamma=0$), we observe that $g_{\rm cr}$ decreases monotonously as $\alpha_R$ increases (see Fig.~\ref{fig:alphagamma0}). The behaviour is the opposite in the presence of  surface tension, where the marginal stability threshold $g_\text{cr}$ monotonously increases with $\alpha_R$. For the range of parameters in which the critical mode displays an interfacial instability, we observe that $g_\text{cr}$ increases linearly with $\alpha_R$. \\
As regards the critical wavenumber $m_{\rm cr}$, we can see that it increases in all the cases, both in the presence and in the absence of surface tension.

We observe that our model captures the main features of organoid development. First, for small values of $\alpha_\gamma$, a morphological transition takes place at the free boundary, giving rise to a wrinkling pattern: as $\alpha_\gamma$ increases, we observe a decrease of the critical wavenumber and a higher marginal stability threshold. 
This is in agreement with the experiments of \cite{Karzbrun_2018}: if the cells have the LIS1 $+/-$ mutation, the authors measured that the elastic modulus of the cells is $2.7$ times lower than the one of healthy cells. In our model this reduction of the stiffness is equivalent to an increase of $\alpha_\gamma$. They have also reported a reduction of the number of folds in organoids affected by lissencephaly. As one can observe from the plots of Figs.~\ref{fig:fixalphaR}-\ref{fig:fixalphagamma} and from the morphological diagram of Table~\ref{tab:parametricplot}, as we increase $\alpha_\gamma$ the number of wrinkles decreases and the critical threshold increases, in accordance with \cite{Karzbrun_2018}'s experiments (these results are shown in \cite{Karzbrun_2018}, Figure 5). Furthermore, for large values of $\alpha_\gamma$, wrinkles at the free surface are completely absent, as happens in the most serious case of lissencephaly.

Compared to the model proposed by \cite{balbi2018mechanics}, in which the authors do not take into account the tissue surface tension, our theoretical description presents some advantages. In fact, we do not introduce different shear moduli for the cortex and the lumen to modulate the critical wavenumber and the critical growth threshold. This choice is motivated by the experimental results of \cite{Karzbrun_2018}: the authors reported a unique value of elastic modulus for the organoid and they did not experimentally measure a change in the shear moduli between the cortex and the lumen.\\
In our model the selection of the critical wavenumber is controlled by the competition between surface capillary energy and bulk elasticity.
Furthermore, we are able to justify the complete absence of surface wrinkles in the most severe cases of lissencephaly, which corresponds to the case of large $\alpha_\gamma$: large values of $\alpha_\gamma$ corresponds to both a high value of surface tension and a very soft material, the shear modulus is decreasing.

Notwithstanding the good agreement with the experimental results of our model, it is to be reported that models based on solid mechanics of brain organoids have been recently criticized by \cite{Engstrom_2018}. The authors observe that the folds of the  cortex display an ``antiwrinkling'' behaviour: the cortex is thicker in correspondence of furrows and thinner at the ridges of wrinkles. The authors claim that solid models do not show this feature and, thus, they are inadequate to model multicellular aggregates. In the next section we implement a numerical code to approximate the fully non-linear problem and we show that the ``antiwrinkling'' phenomenon is provoked by tissue surface tension.

\section{Post-buckling analysis}
\label{sec:post_buckling}

\subsection{Description of the numerical method}

\begin{figure}[b!]
\centering
\includegraphics[width=0.6\textwidth]{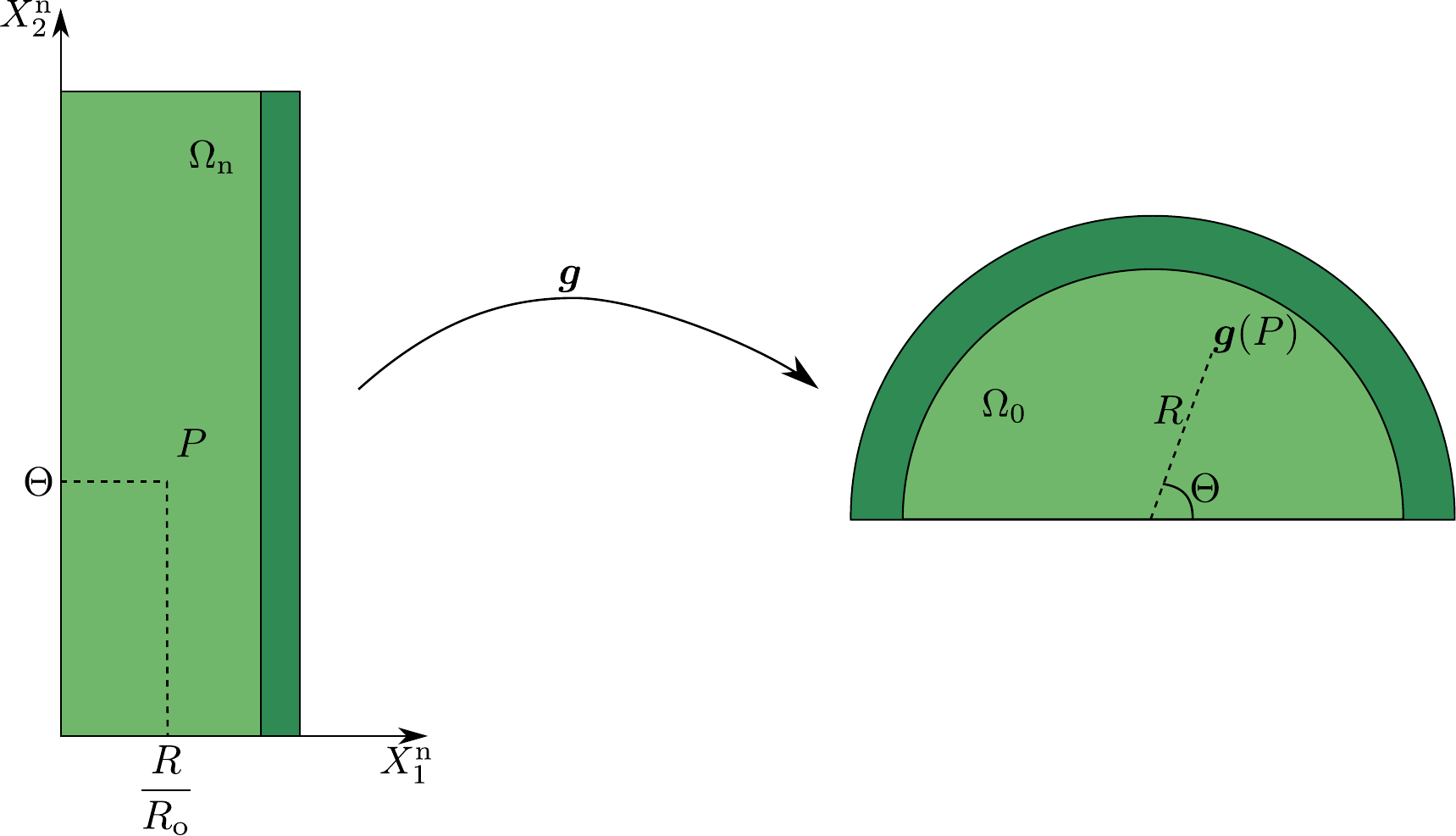}
\caption{Representation of the conformal mapping $\vect{g}$ that maps the computational domain $\Omega_\text{n}$ to the reference configuration $\Omega_0$}
\label{fig:comp_domain}
\end{figure}

In this section we show the results of the numerical approximation of the non-linear problem given by Eqs.~\eqref{eq:balance}-\eqref{eq:BCcenter}-\eqref{eq:PN} to investigate the post-buckling behaviour of the organoid.
We use as computational domain the rectangle obtained through the conformal mapping corresponding to the polar coordinate transformation, as done in \citep{Riccobelli_2017}: let
\[
\Omega_\text{n} = (0,\,1)\times(0,\,\pi).
\]
Given $\vect{X}_\text{n}\in\Omega_\text{n}$, the components represent the referential radial coordinate normalized with respect to the external radius, and the referential polar angle, respectively:
\[
\left\{
\begin{aligned}
&X^\text{n}_1 = \frac{R}{R_\text{o}},\\
&X^\text{n}_2 = \Theta,
\end{aligned}
\right.
\]
as represented in Fig.~\ref{fig:comp_domain}. The function
\[
\vect{g}(\vect{X}_\text{n}) = \left[R_\text{o} X^\text{n}_1\cos(X^\text{n}_2), R_\text{o} X^\text{n}_1\sin(X^\text{n}_2)\right]
\]
maps the computational domain to a half circle, which represent half of the reference configuration. The full domain can be obtained thanks to the axial symmetry of the problem.

\begin{figure}[t!]
\centering
\includegraphics[width=0.5\textwidth]{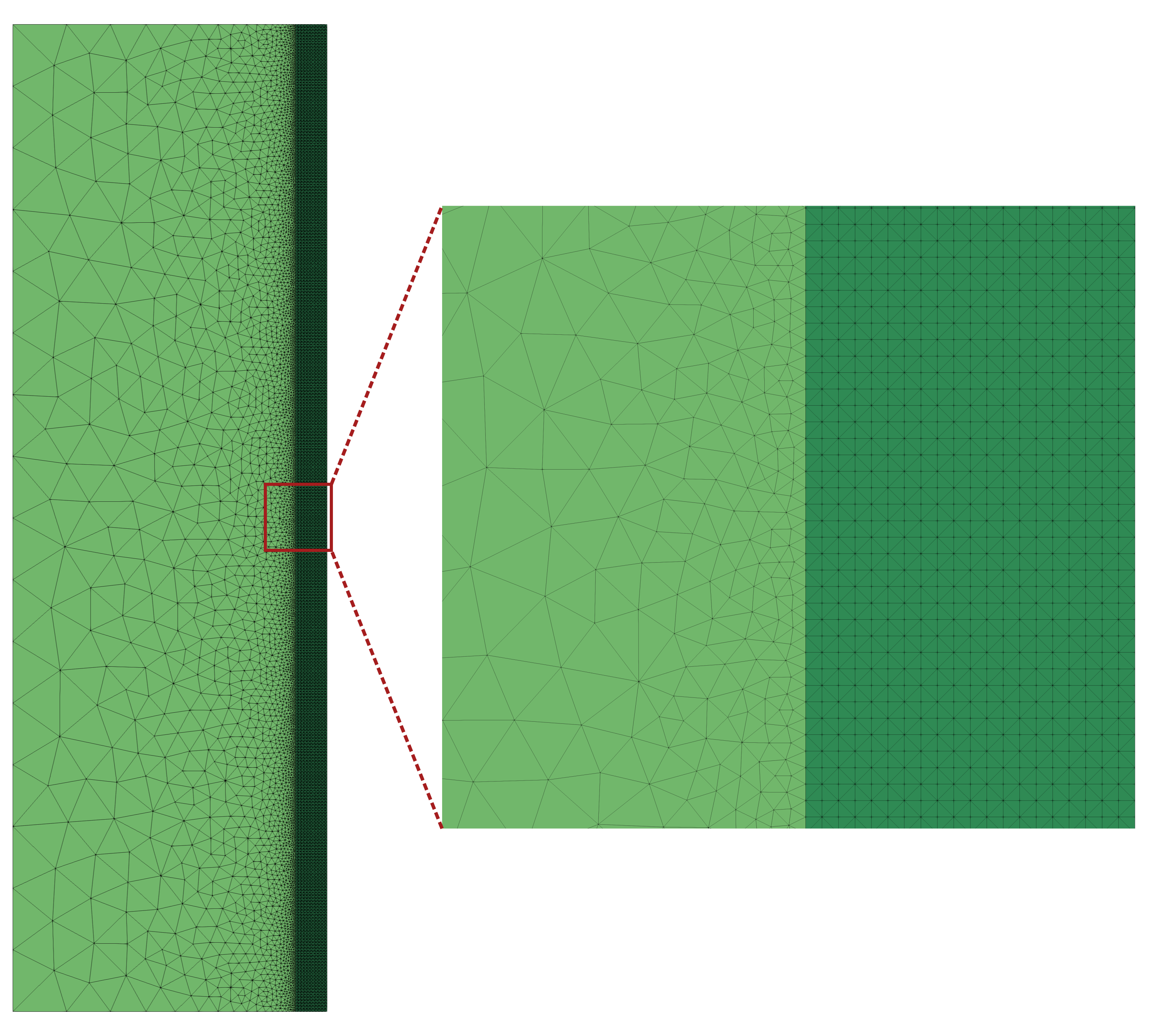}
\caption{Mesh generated through GMSH for $\alpha_R=0.9$. The maximum diameter of this mesh elements is $0.2488$ while the minimum diameter is $0.0017$.}
\label{fig:mesh}
\end{figure}

We discretise the computational domain through the software GMSH 
\citep{Geuzaine_2009}. We use a triangular grid, with a progressive refinement of the elements from $X_1^\text{n} = 0$ up to $X_1^\text{n} = \alpha_R$. In the cortex we instead use a structured mesh (i.e. for $\alpha_R<X_1^\text{n}<1$), see Fig.~\ref{fig:mesh}.

We implement a numerical code based on the mixed finite element method to enforce the incompressibility constraint \eqref{eq:incompressibility}. We discretise the displacement field $\vect{u}$ using continuous, piecewise quadratic functions, while we approximate the pressure through piecewise constant functions. The corresponding mixed finite element is the  $\vect{P}_2-P_0$ element, which is numerically stable for the incompressible hyperelastic problem \citep{Boffi_2013}.
We use an index $h$ when we refer to the discretised counterpart of the mathematical quantities. We adimensionalise the system of equation with respect to $\mu$ and $R_\text{o}$ as we did at the continuum level. We set the following discrete boundary conditions
\[
\left\{
\begin{aligned}
&\vect{u}_h = \vect{0} &&\text{if }X^\text{n}_1 = 0,\\
&\vect{u}_h\cdot\vect{e}_2 = 0 &&\text{if }X^\text{n}_2 = 0\text{ or }X^\text{n}_2 = \pi,\\
&\tens{P}_h^T\vect{e}_2 \cdot\vect{e}_1 = 0 &&\text{if }X^\text{n}_2 = 0\text{ or }X^\text{n}_2 = \pi,\\
&\tens{P}_h^T\vect{e}_1 = \det{\tens{F}_h}\alpha_\gamma \mathcal{K}_h \tens{F}_h^{-T}\vect{e}_1 &&\text{if }X^\text{n}_1 = 1,\\
\end{aligned}
\right.
\]
where $(\vect{e}_1,\,\vect{e}_2)$ represents the canonical vector basis.

We solve the discretised form of the equilibrium equation \eqref{eq:balance} in the Lagrangian form using a Newton method. The control parameter $g$ is incremented of $\delta g$ when the Newton method converges, the numerical solution is used as initial guess for the following Newton cycle. The increment $\delta g$ is automatically reduced near the theoretical marginal stability threshold and when the Newton method does not converge. The numerical simulation is stopped when $\delta g < 10^{-6}$.
To trigger the mechanical instability, a small perturbation of an amplitude of $\sim 10^{-5}$, having the shape of the critical mode computed in Section~\ref{sec:lin_stab}, is applied at the free boundary of the mesh. We have numerically verified that the wavelength of the buckled pattern is not sensitive to the applied imperfection and the only effect is a slight anticipation of the instability threshold.

The numerical algorithm is implemented in Python through the open-source computing platform FEniCS (version 2018.1) \citep{Logg_2012}. The computation of the weak form and of the Jacobian necessary to solve each step of the Newton method are computed from the total energy through the library UFL \citep{Aln_s_2014}. Surface tension is introduced in the numerical algorithm following the implementation proposed in \citep{Mora_2013}.
We use PETSc \citep{balay2018petsc} as linear algebra back-end and MUMPS \citep{Amestoy_2000} as linear solver.

\subsection{Results of the finite element simulations}

In this section, we discuss the results of the numerical simulations for $\alpha_R = 0.9$.
\begin{figure}[t!]
	\begin{subfigure}{.5\linewidth}
		\centering
		\includegraphics[width=1\textwidth]{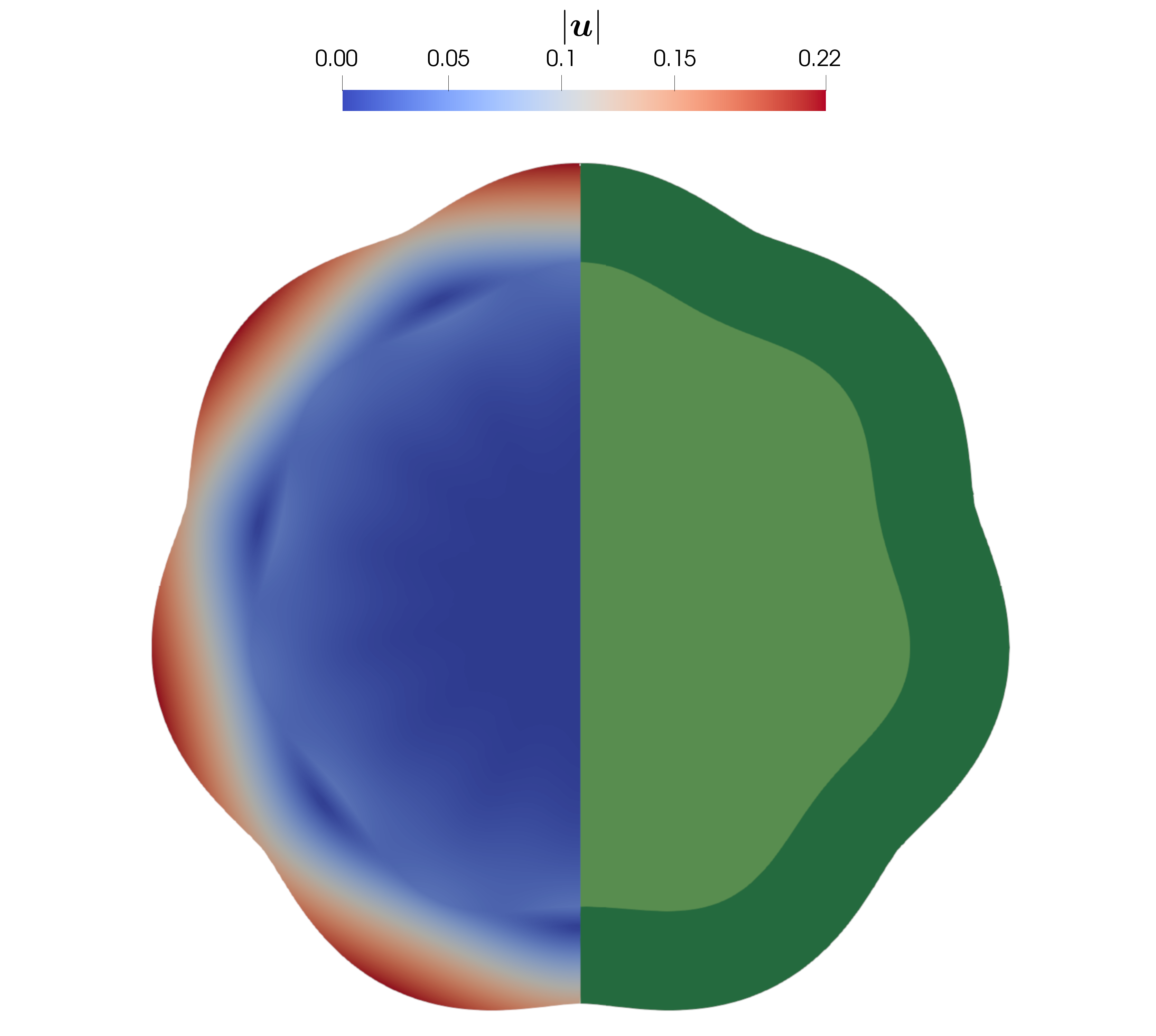}
		\caption{$\alpha_\gamma = 0$}
		\label{fig:deformazionealphagamma0}
	\end{subfigure}%
	\begin{subfigure}{.5\linewidth}
		\centering
		\includegraphics[width=1\textwidth]{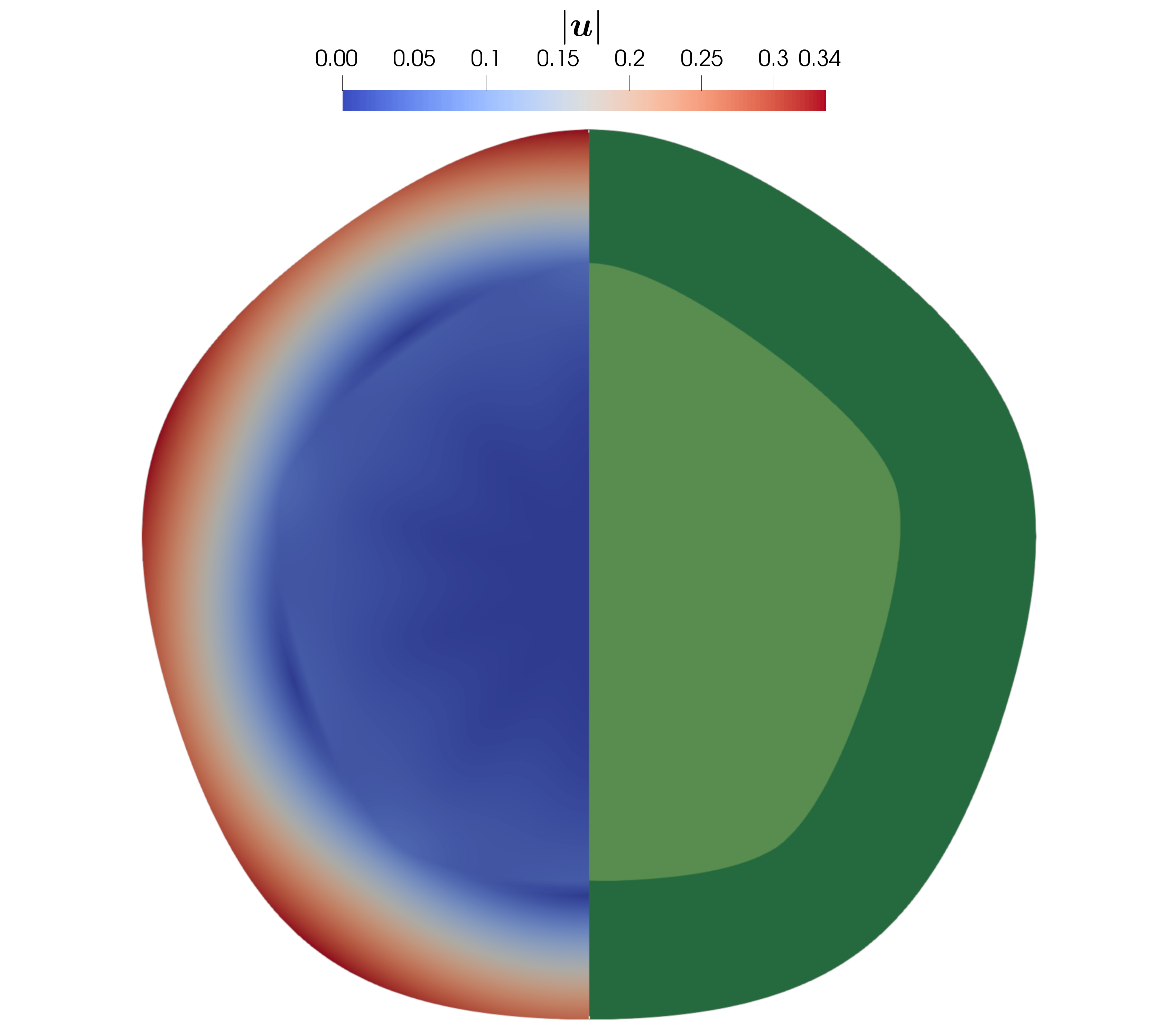}
		\caption{$\alpha_\gamma = 0.5$}
		\label{fig:deformazionealphagamma05}
	\end{subfigure}
	\caption{Buckled configuration for (a) $\alpha_\gamma=0$ and $g=1.7556$ and (b) $\alpha_\gamma=0.5$ and $g = 2.1646$.}
	\label{fig:deformazione}
\end{figure}
In Fig.~\ref{fig:deformazione}, we plot the buckled configuration of the organoid. As predicted by the critical modes of the linear stability analysis plotted in Table~\ref{tab:parametricplot}, in presence of surface tension (Fig.~\ref{fig:deformazionealphagamma05}) the free boundary is smoother and rounded.

We define $\Delta r$ as the amplitude of the pattern at the free surface
\[
\Delta r = \max_{\Theta\in[0,\pi]} r(R_\text{o},\,\Theta) - \min_{\Theta\in[0,\pi]} r(R_\text{o},\,\Theta) 
\]
where $r$ denotes the actual radial position of the point with polar coordinates $(R,\,\Theta)$.

\begin{figure}[t!]
	\begin{subfigure}{.5\linewidth}
		\centering
		\includegraphics[width=1\textwidth]{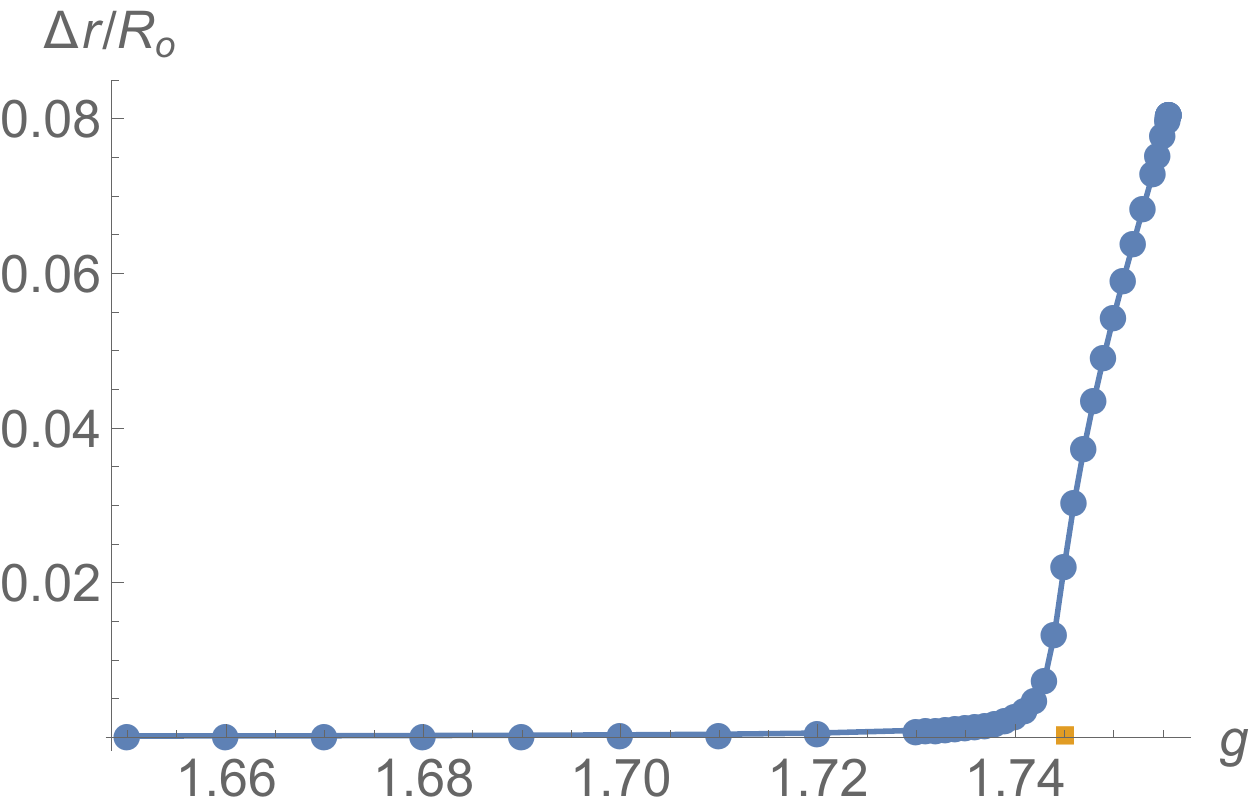}
		\caption{$\alpha_\gamma = 0$}
		\label{fig:ampiezzaalphagamma0}
	\end{subfigure}%
	\begin{subfigure}{.5\linewidth}
		\centering
		\includegraphics[width=1\textwidth]{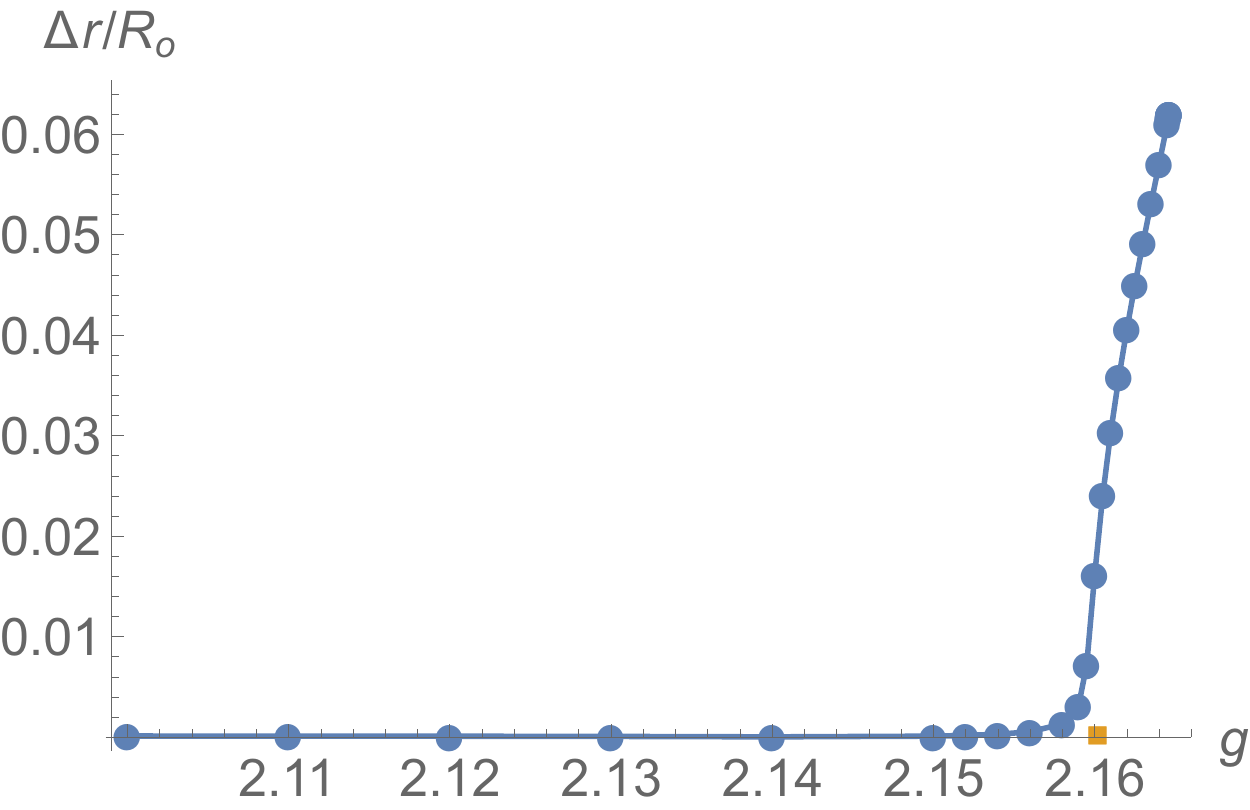}
		\caption{$\alpha_\gamma = 0.5$}
		\label{fig:ampiezzaalphagamma05}
	\end{subfigure}
	\caption{Bifurcation diagrams for (a) $\alpha_\gamma=0$ and (b) $\alpha_\gamma=0.5$. The orange square denotes the theoretical marginal stability threshold computed as exposed in Section~\ref{sec:lin_stab}. The good agreement between the linear stability analysis and the finite element code outcomes validates the numerical algorithm.}
	\label{fig:ampiezza}
\end{figure}

In Fig.~\ref{fig:ampiezza}, we show how the amplitude of the pattern $\Delta r$  evolves with respect to the growth rate $g$. We observe that there is an excellent agreement with the marginal stability thresholds computed in the previous section, proving the consistence of the numerical code with respect to the theoretical predictions.
Both the bifurcation diagrams exhibit a continuous transition from the unbuckled to the buckled configuration, displaying the typical behaviour of a supercritical pitchfork bifurcation. Let
\[
E = \int_\Omega \psi(\tens{F})\,dV + \gamma \int_{\partial \Omega} |\tens{F}^{-T}\vect{N}|\,dS
\]
be the total mechanical energy of the organoid. We compute the ratio of the energy $E_\text{th}$ of the base solution given by Eqs.~\eqref{eq:r<ri}-\eqref{eq:ri<r<ro} and the energy $E_\text{num}$ arising from the numerical simulations.

\begin{figure}[t!]
	\begin{subfigure}{.5\linewidth}
		\centering
		\includegraphics[width=1\textwidth]{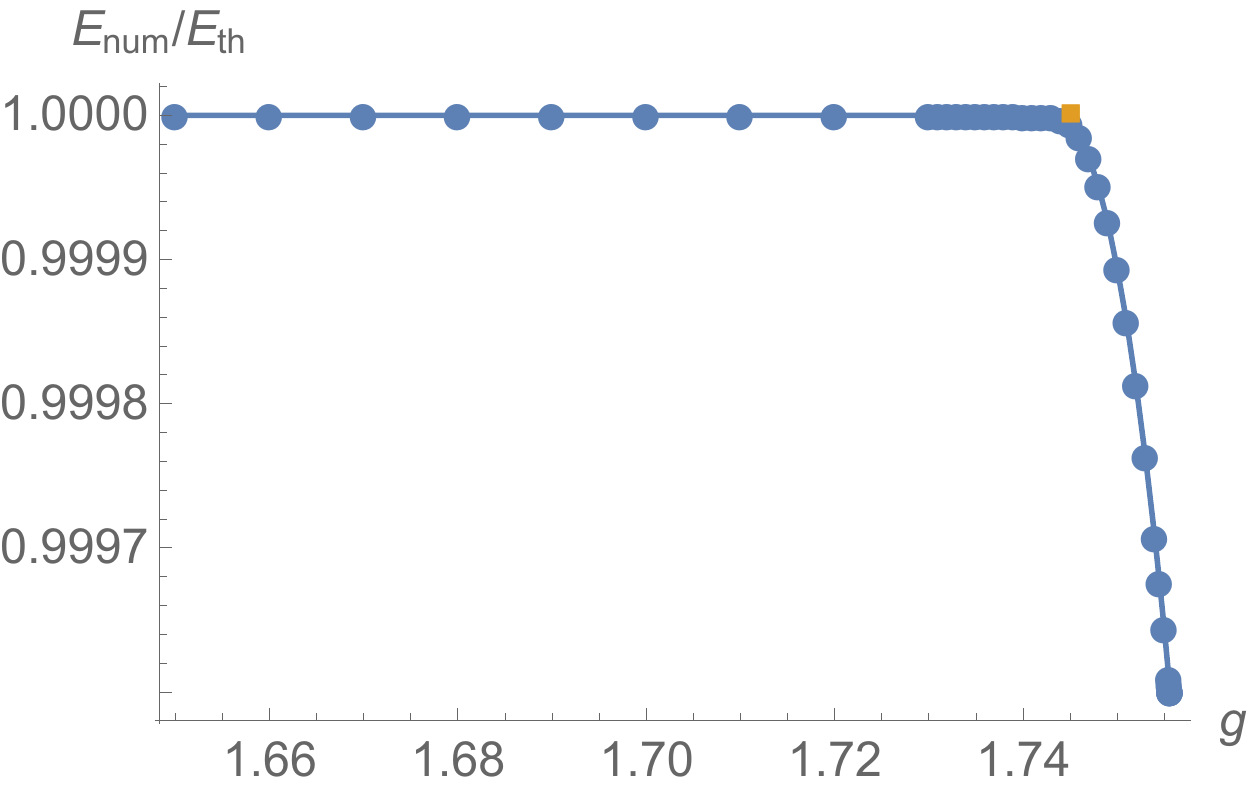}
		\caption{$\alpha_\gamma = 0$}
		\label{fig:energiaalphagamma0}
	\end{subfigure}%
	\begin{subfigure}{.5\linewidth}
		\centering
		\includegraphics[width=1\textwidth]{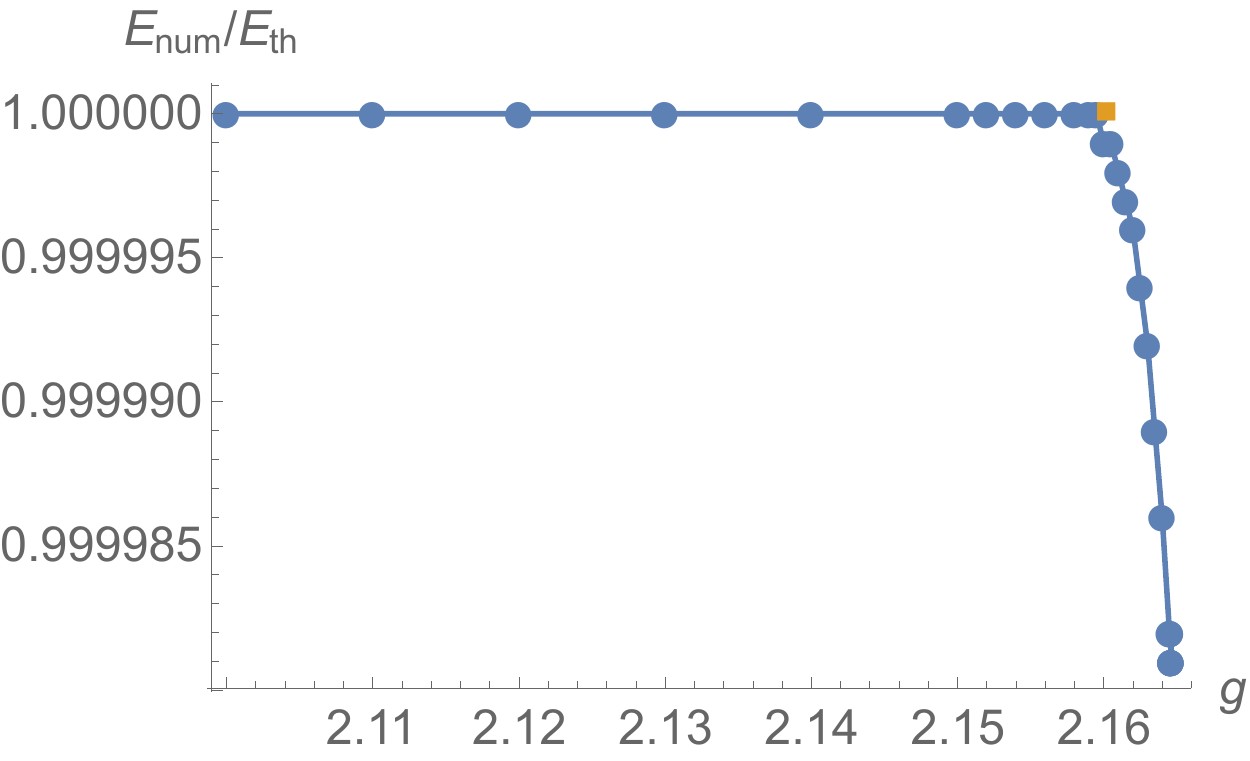}
		\caption{$\alpha_\gamma = 0.5$}
		\label{fig:energiaalphagamma05}
	\end{subfigure}
	\caption{Plot of the ratio $E_\text{th}/E_\text{num}$ for (a) $\alpha_\gamma=0$ and (b) $\alpha_\gamma=0.5$. The orange square denotes the theoretical marginal stability threshold computed as exposed in Section~\ref{sec:lin_stab}.}
	\label{fig:energia}
\end{figure}

In Fig.~\ref{fig:energia} we plot the ratio $E_\text{num}/E_\text{th}$ versus the control parameter $g$ for both $\alpha_\gamma=0$ and $\alpha_\gamma = 0.5$. In both cases, the buckled configuration exhibits a total lower mechanical energy with respect to the unbuckled state. Furthermore we observe that the energy lowers continuously, confirming that the bifurcation is supercritical. 
\begin{figure}[t!]
	\begin{subfigure}{.5\linewidth}
		\centering
		\includegraphics[width=1\textwidth]{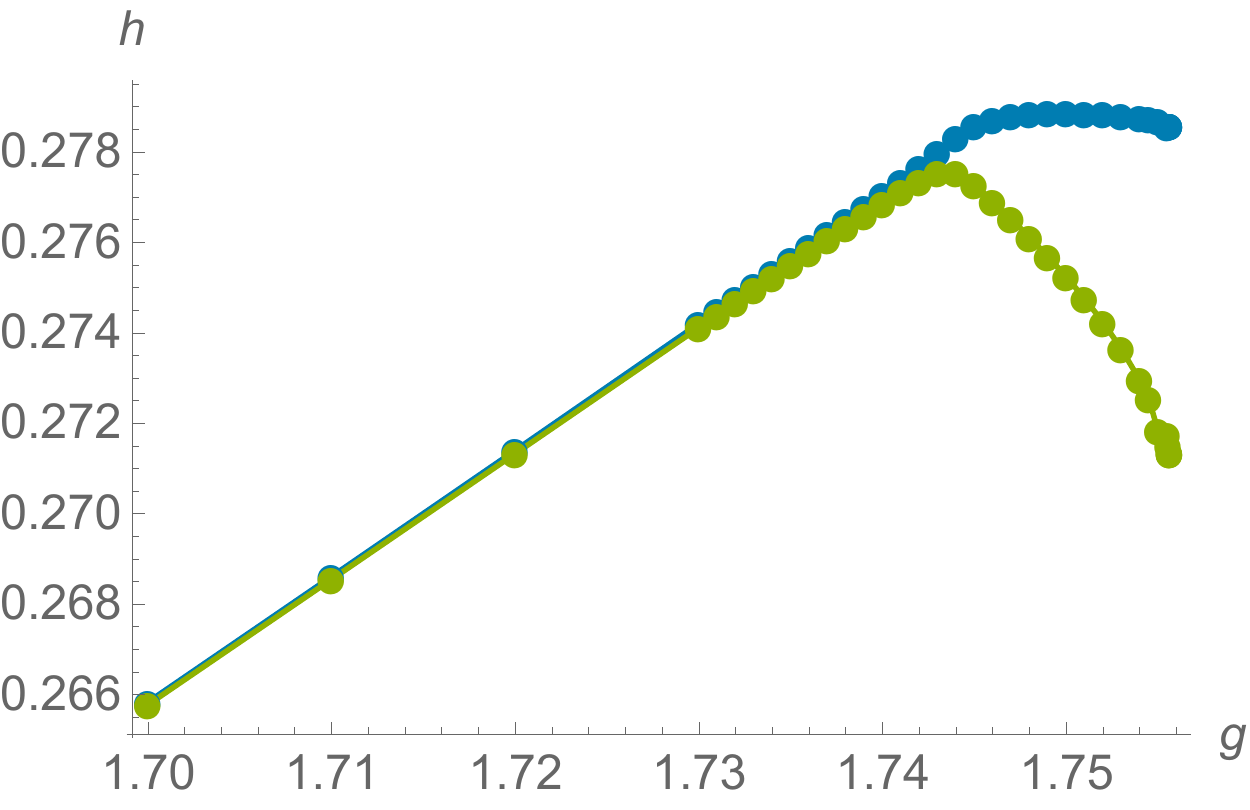}
		\caption{$\alpha_\gamma = 0$}
		\label{fig:spesssorealphagamma0}
	\end{subfigure}%
	\begin{subfigure}{.5\linewidth}
		\centering
		\includegraphics[width=1\textwidth]{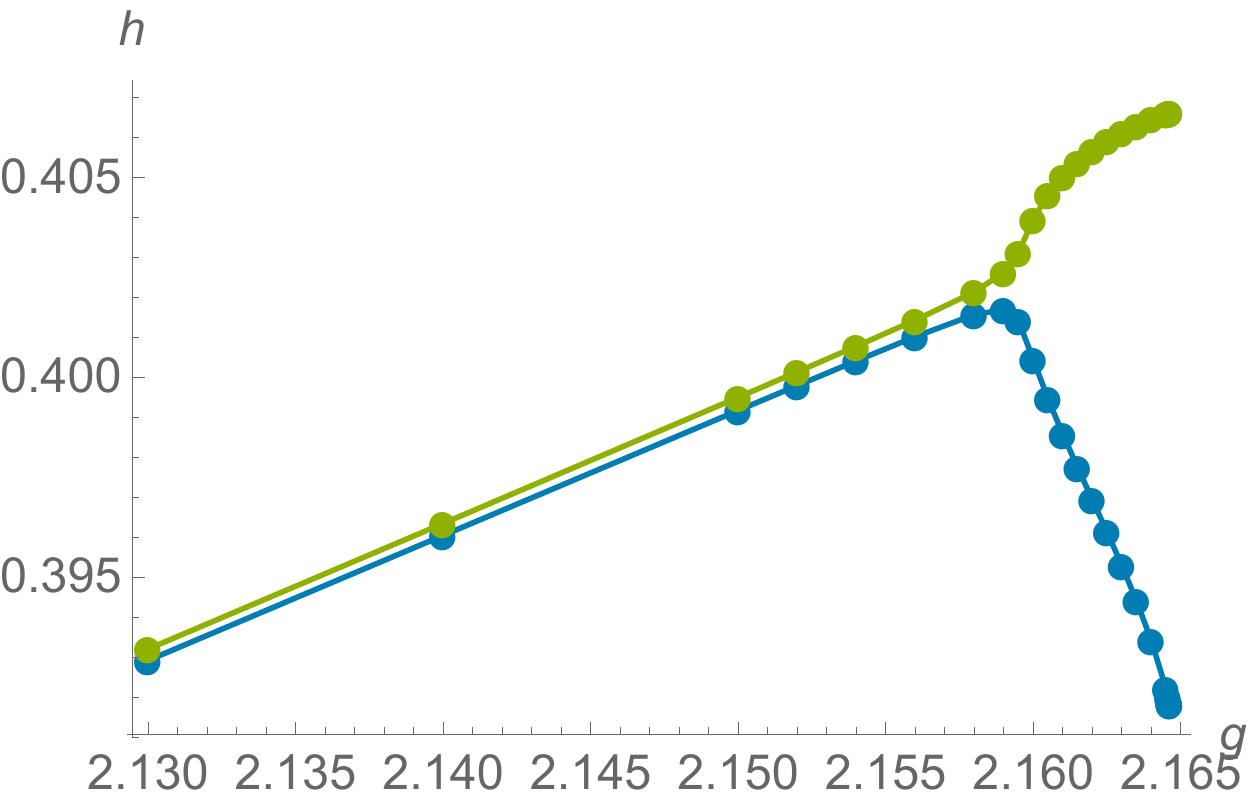}
		\caption{$\alpha_\gamma = 0.5$}
		\label{fig:spesssorealphagamma05}
	\end{subfigure}
	\caption{Thickness of the cortex of ridges (blue) and furrows (green) for (a) $\alpha_\gamma=0$ and (b) $\alpha_\gamma=0.5$. The latter situation, in which the thickness of ridges is higher that the one of furrows, corresponds to the ``antiwrinkling'' behaviour described in \citep{Engstrom_2018}.}
	\label{fig:thick_cortex}
\end{figure}

Finally, we compute the thickness of the cortex at the ridges and at the furrows of the buckled configuration (which correspond to the gyri and the sulci of the fully developed brain respectively).
We observe that the thickness of the ridges is higher than the one of the furrows for $\alpha_\gamma = 0$ (see Fig.~\ref{fig:spesssorealphagamma0}), while the behaviour is the opposite in the case of $\alpha_\gamma = 0.5$ (see Fig.~\ref{fig:spesssorealphagamma05}). The latter case is typical of brain organoids, as observed in \citep{Engstrom_2018}.

These numerical results suggest that the ``antiwrinkling'' behaviour, as named by \cite{Engstrom_2018}, is due to the presence of surface tension, which is highly relevant due to the small radius of a brain organoid.
If we consider a fully developed brain, its typical size is of order of decimeters. In this case, keeping $\gamma$ and $\mu$ fixed, $\alpha_\gamma$ is reduced by five order of magnitudes with respect to the case of the organoid and the contribution of surface tension to the total energy becomes negligible. Experimental results show that gyri are thicker than sulci \citep{Holland_2018}, differently from what happens at the small length scales of the brain organoid. This is in agreement with the outcomes of our model in the case $\alpha_\gamma=0$, confirming that the ``antiwrinkling'' behaviour is caused by the competition of bulk elastic energy and the surface tension.

However, our numerical algorithm has some limitations. The Newton method does not converge anymore slightly past the marginal stability threshold. In fact, the bifurcation diagrams of Fig.~\ref{fig:ampiezza} show that the amplitude of the pattern increases very rapidly beyond the marginal stability threshold. As one can observe in Fig.~\ref{fig:deformazionealphagamma0}, the deformation tends to localize near the furrows, highlighting a possible subcritical transition to a folded state, which would be in agreement with the experimental observations of \cite{Karzbrun_2018}. The numerical approximation of a layer growing on a substrate with similar mechanical properties is particularly tricky, even numerically~\citep{Fu_2015}. Future efforts will be devoted to the improvement of the numerical scheme, implementing an arclength continuation method to study the wrinkle-to-fold transition in brain organoids.

\section{Discussion and concluding remarks}
\label{sec:discussion}
In this work, we have developed a model of brain organoids to describe the formation of cerebral sulci and to investigate the influence of surface tension on such a morphogenetic process. In Section~\ref{sec:surface_tension} we have computed that the tissue surface tension acting on a solid multicellular spheroid is $\gamma = 10^{-1} \, \mathrm{N}/\mathrm{m}$, using the data reported in \cite{Lee_2019}. We exploit this measure as a qualitative estimate of tissue surface tension of a generic cellular aggregate. In fact, measures performed on different embryonic tissues modelled as fluids all have the same order of magnitude \citep{Sch_tz_2008}. Thus, we do not expect that the tissue surface tension acting on brain organoids dramatically differs from the one acting on multicellular spheroids.

Then, we have built in Section~\ref{sec:elastic_model} an elastic model of brain organoids. They are described as disks surrounded by a growing rim and subjected to a surface tension generated by intercellular adhesion forces \citep{Manning_2010}. We have assumed that the two regions (disk and outer rim) are composed of the same incompressible neo-Hookean material. The system is governed by the dimensionless parameters $g$, i.e.~the growth rate of the cortex with respect to the lumen, $\alpha_R$ and $\alpha_\gamma$, which are the lumen radius and the capillary length, normalized with respect to the initial radius of the organoid, respectively.

We have computed a radially symmetric solution and we have studied its linear stability in Section~\ref{sec:lin_stab} using the theory of incremental deformations. We have rewritten the linear stability analysis into an optimal Hamiltonian system using the Stroh formulation \citep{Fu_2007}. The impedance matrix method is adapted to take into account the boundary contribution of surface tension.

The outcomes are discussed in Section~\ref{subsec:result_linear}.
The introduction of a tissue surface tension of the same order as the estimate computed in Section~\ref{sec:surf_tens_spheroid} strongly influence the stability of the base solution and alters the critical wavelength. This result suggests that the interplay between elasticity and tissue surface tension plays a crucial role in controlling pattern selection in embryo morphogenesis.
The predicted critical wavelength is always finite (see Fig.~\ref{fig:alphagamma0}). In our model the softening of cells, due to lissencephaly, corresponds to an increase of the parameter $\alpha_\gamma$, strengthening the role of tissue surface tension and leading to a lower critical wavenumber. Experimental observations of \cite{Karzbrun_2018} report a similar behaviour: when the LIS1 $+/-$ mutation is present, the shear modulus of the organoid is reduced and the wavenumber decreases compared with the healthy organoids. This suggest that the reduction of brain sulci due to lissencephaly is due to the competition between elastic and surface energies in early embryogenesis.
Moreover, we have observed that, for larger $\alpha_\gamma$, a transition from a surface to an interfacial instability occurs: buckling localizes at the interface between the cortex and the lumen. In this case the cortex remains smooth as one can observe in the most severe cases of lissencephaly. The results are reported in Figs.~\ref{fig:fixalphaR}-\ref{fig:agcr_vs_ar}-\ref{fig:fixalphagamma} and in Tab.~\ref{tab:parametricplot}.

Finally, in Section~\ref{sec:post_buckling}, we have implemented a finite element code to approximate the fully non-linear problem. The algorithm is based on a mixed variational formulation and the Newton method. The outcomes of the numerical simulations are reported in Fig.~\ref{fig:deformazione}-\ref{fig:thick_cortex}. These results show that tissue surface tension rounds the external boundary. Both in the presence and in the absence of surface energy, the bifurcation is supercritical, displaying a continuous transition from the unbuckled to the buckled state. Contrarily to the bilayer model without surface tension, our model reproduces ``antiwrinkling'' behaviour of the cortex (namely the thickness of the outer layer is larger in the furrows). This suggests that this unconventional variation of cortex thickness is due to tissue surface energy. This strengthens the importance of considering surface tension in the modelling of cellular aggregates.

Summing up, our model suggests a possible purely mechanical explanation of a number of open questions:
\begin{itemize}
\item brain sulci in organoids are generated by a mechanical instability triggered by cortex growth;
\item the ``anti-wrinkling'' behaviour is induced by the presence of surface tension which tends to reduce the perimeter of the organoid as proved by the numerical simulations;
\item the reduction of stiffness of the cell decreases the role of elasticity and enhances the effect of surface tension (leading to an increased $\alpha_\gamma$ while keeping $\gamma$ constant). A reduction of stiffness of the organoid has different consequences: the onset of bifurcation is delayed, the critical wavelength increases, the cortex is thicker and, in the most severe cases, the surface instability is absent.
\end{itemize}

This work opens the path to other studies, such as the analysis of the effect of surface tension on the growth of embryos and tumour spheroid. Our paper proposes a theoretical explanation of purely mechanical nature of known experiments; of course further investigations will be needed to validate it. What emerges from our model is that tissue surface tension cannot be disregarded whenever a living tissue is characterized by relatively small length scales or by small elastic moduli.
 
Future efforts will include the improvement of the numerical scheme in order to capture possible secondary bifurcations. Possible extensions include the implementation of an arclength continuation method to improve the numerical convergence in presence of turning points. From an experimental point of view, it would be important to quantitatively measure the surface tension acting on the organoid cortex. Another possible line of research is the study of the influence of surface tension on the growth of cellular aggregates.

\section*{Acknowledgments}
\noindent The authors wish to thank D. Ambrosi, P. Ciarletta, A. Marzocchi and G. Noselli for helpful suggestions and fruitful discussions. DR gratefully acknowledge the support of the European Research Council (AdG-340685 MicroMotility).

\begin{appendices}
\section{Computation of the incremental curvature}
\label{app:A}
The external boundary in the actual incremental configuration is given by
\begin{equation}
\label{eq:ext_bound}
\vect{\alpha}(\theta) = (r_\text{o}+u(r_\text{o},\,\theta))\left[\cos(\theta+v(r_\text{o},\,\theta)),\,\sin\left(-\theta-v(r_\text{o},\,\theta\right)\right],
\end{equation}
since the parametrization of $\vect{\alpha}$ is counter-clockwise \eqref{eq:clockwise}. By differential geometry, the oriented curvature of $\vect{\alpha}$ is given by
\begin{equation}
\label{eq:curvatura_formula}
\mathcal{K} = \frac{\alpha_x'(\theta)\alpha''_y(\theta)-\alpha_y'(\theta)\alpha''_x(\theta)}{|\vect\alpha'(\theta)|^3},
\end{equation}
where with $'$ we denote the derivative with respect to $\theta$. By combining Eq.~\eqref{eq:ext_bound} and Eq.~\eqref{eq:curvatura_formula} we obtain
\begin{equation}
\label{eq:curvatura_tot}
\mathcal{K}=\frac{\left(v'+1\right) \left((u+r_\text{o}) \left(u''-(u+r_\text{o}) \left(v'+1\right)^2\right)-2 u'^2\right)-(u+r_\text{o}) u' v''}{\left(u'^2+(u+r_\text{o})^2 \left(v'+1\right)^2\right)^{3/2}}.
\end{equation}
We can linearise the relation~\eqref{eq:curvatura_tot} with respect to $u$, $v$, and their derivatives to get the following expression of the incremental curvature
\[
\delta\mathcal{K} = \frac{u''+u'}{r_\text{o}^2}.
\]

\section{Role of the Matrigel embedment}
\label{app:B}

\begin{figure}[t!]
\centering
\includegraphics[width=0.5\textwidth]{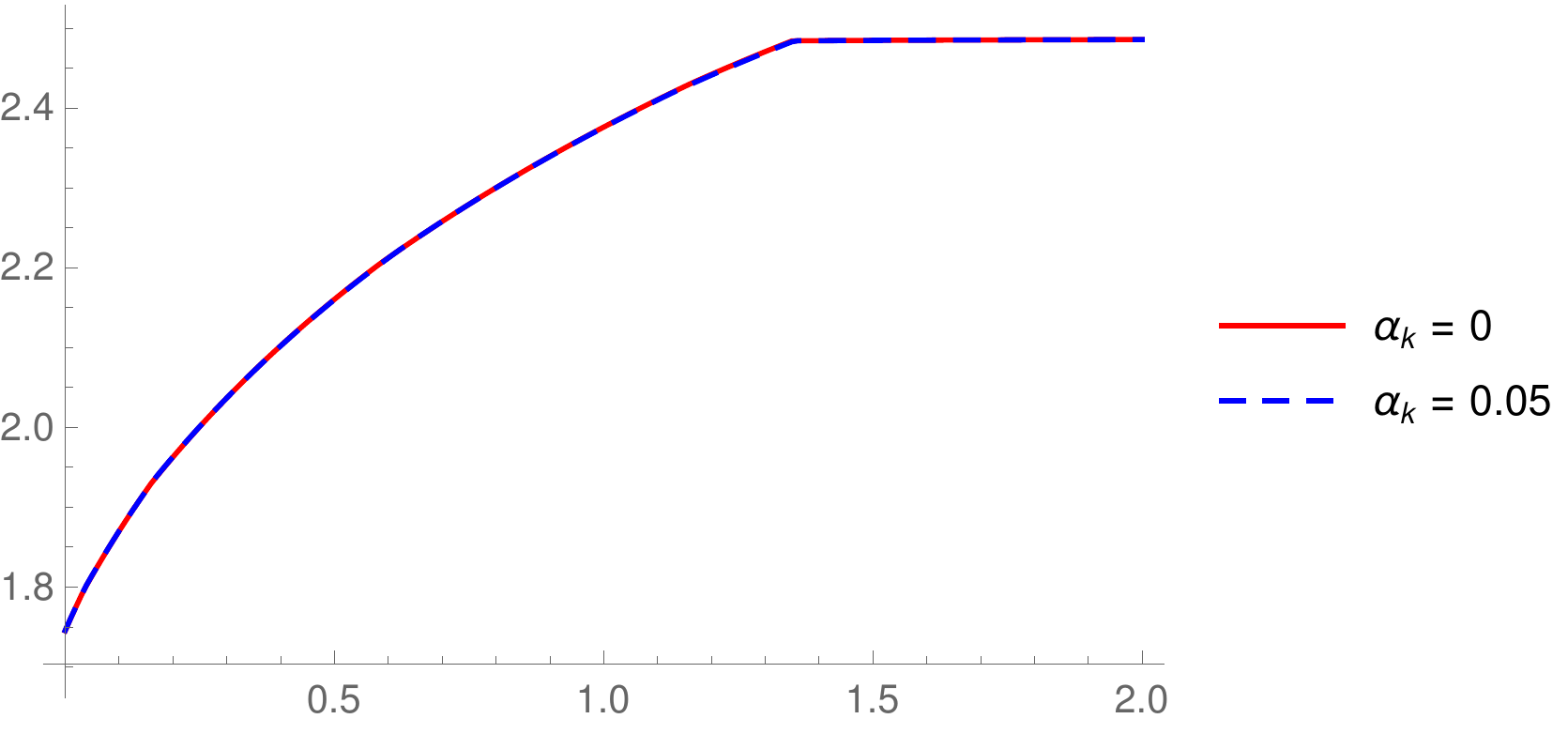}\includegraphics[width=0.5\textwidth]{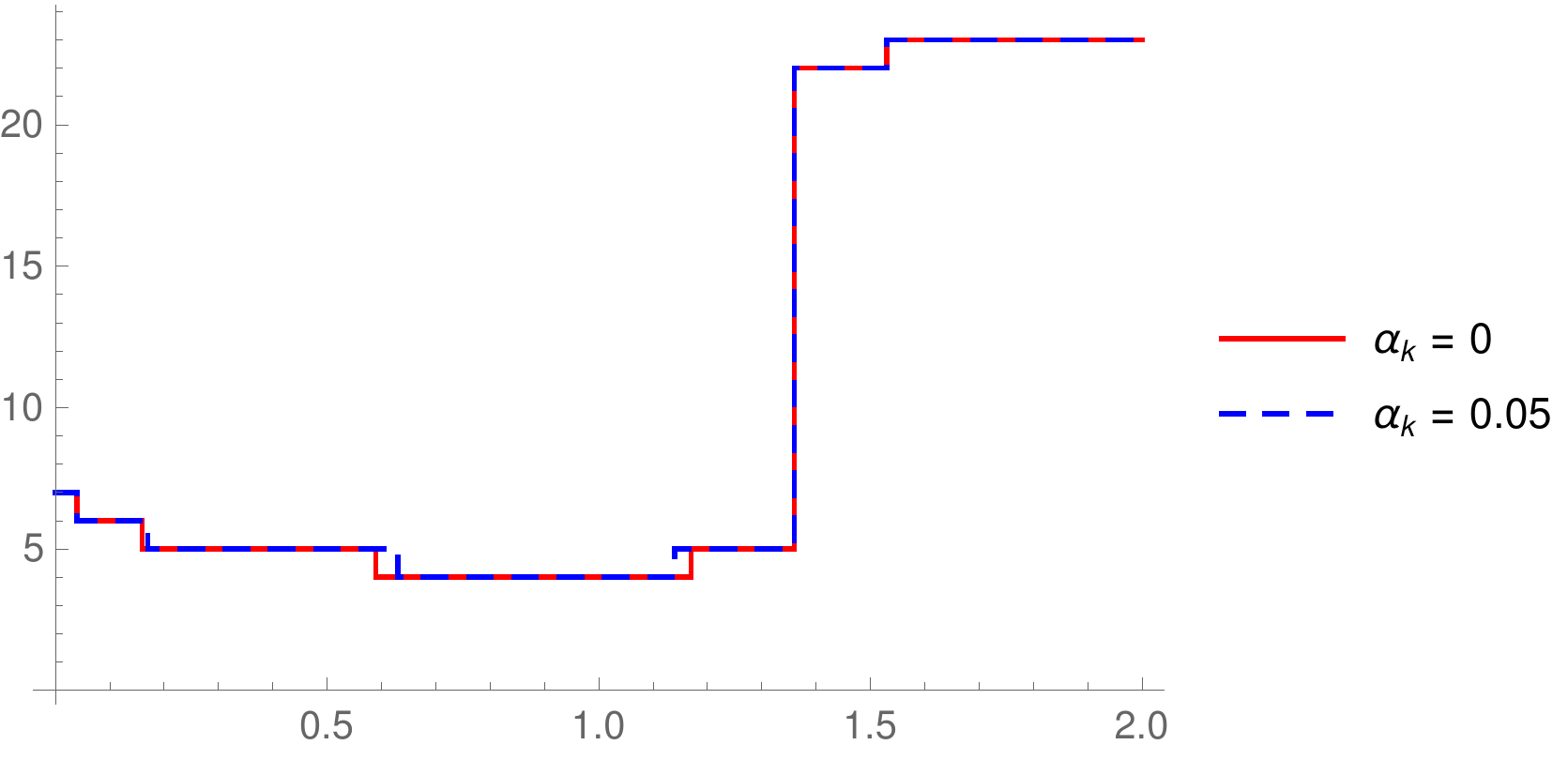}
\caption{Marginal stability threshold $g_\mathrm{cr}$ (top) and critical wavenumber $m_\mathrm{cr}$ (bottom) versus $\alpha_\gamma$ for $\alpha_R = 0.9$ and $\alpha_k = 0,\,0.5$.}
\label{fig:confronto}
\end{figure}
Let $k_s$ be the elastic constant of the springs per unit length. Let $\mu_M\simeq\mu$ be the shear modulus of the Matrigel.
 The boundary condition Eq.~\eqref{eq:PN} modifies into
\[
\tens{P}^T\vect{N} = (\det\tens{F})\gamma\mathcal{K}\tens{F}^{-T}\vect{N} - k_\mathrm{s}\vect{u}.
\]
The radially symmetric solution given by Eqs.~\eqref{eq:r<ri}-\eqref{eq:ri<r<ro} is still a solution of the modified problem, the only difference is that the pressure increases of a constant $k_\mathrm{s}R_o(1-R_o/r_o)$. Denoting by $R_M$ the characteristic distance among the organoids, from a dimensional evaluation we obtain
\[
k_s \simeq \frac{\mu}{R_M}.
\]
We adimensionalize $k_s$ with respect to $\mu$ and $R_\mathrm{o}$, obtaining
\[
\alpha_k = \frac{k_s R_o}{\mu}\simeq\frac{R_o}{R_M}\simeq \frac{500\mu\mathrm{m}}{1 \mathrm{cm}} = 0.05.
\]
From this computation it is likely that the force exerted by the Matrigel on the organoid is negligible. To verify it, we have modified the linear stability of exposed in Section \ref{sec:lin_stab} to account for the Matrigel embedment.

Following the procedure exposed in \citep{Riccobelli_2018}, we can compute the marginal stability threshold using the same algorithm exposed in the manuscript by just modifying Eq.~\eqref{eq:auxiliary_condition} into 
\[
\tens{Z}_\mathrm{o} = -\frac{\alpha_\gamma}{r_\text{o}}\begin{bmatrix}
m^2 &m\\
m &1
\end{bmatrix} - \alpha_k\tens{I}.
\]

The marginal stability threshold are only slightly influenced by the presence of the linear springs at the boundary, as shown by Fig.~\ref{fig:confronto}.
From these results, it is clear that the presence of Matrigel has very little influence on the behaviour of the organoid, as observed in \citep{Karzbrun_2018}.
\end{appendices}

\bibliographystyle{abbrvnat}
\bibliography{refs}

\end{document}

%% file: defs.tex
\theoremstyle{plain} 
\newtheorem{thm}{Theorem}

\newtheorem{rem}[thm]{Remark}

\DeclareMathOperator{\diver}{div}
\DeclareMathOperator{\grad}{grad}
\DeclareMathOperator{\Diver}{Div}
\DeclareMathOperator{\Grad}{Grad}

\DeclareMathOperator{\tr}{tr}

\DeclareMathOperator{\diag}{diag}

\usepackage{mathtools}
\usepackage{pgfplots}
\pgfplotsset{/pgf/number format/use comma,compat=newest}

\renewcommand\epsilon{\varepsilon}
\newcommand{\R}{\mathbb{R}}

\newcommand{\Cped}{{\rm C}}
\newcommand{\Lped}{{\rm L}}

\newcommand{\vect}[1]{\boldsymbol{#1}}
\newcommand{\tens}[1]{\mathsf{#1}}